\documentclass[useAMS,usenatbib]{mn2e}
\usepackage{graphicx}

\def\solmas{$\mathrm{M_\odot}$\,}
\def\solmasp{$\mathrm{M_\odot}$}

\title[Molecular cloud formation timescales]{How long does it take to form a molecular cloud?}
\author[Clark et al.]{Paul~C.~Clark$^1$, Simon~C.~O.~Glover$^1$, Ralf~S.~Klessen$^1$ \& Ian~A.~Bonnell$^2$\\
$^1$Institut f\"ur Theoretische Astrophysik, Zentrum f\"ur Astronomie der Universit\"at Heidelberg, Albert-Ueberle-Str.\ 2, 69120 Heidelberg \\
$^2$ Scottish Universities Physics Alliance (SUPA), School of Physics and Astronomy,  University of St Andrews, \\ The North Haugh, St Andrews, Fife, KY16 9SS \\
 {\tt email:} p.clark@uni-heidelberg.de, glover@uni-heidelberg.de, klessen@uni-heidelberg.de, iab1@st-and.ac.uk}

\begin{document}

\maketitle

\begin{abstract}
We present the first numerical simulations that self-consistently follow the formation of dense molecular clouds in colliding flows. Our calculations include a time-dependent model for the H$_2$ and CO chemistry that runs alongside a detailed treatment of the dominant heating and cooling processes in the ISM. We adopt initial conditions characteristic of the warm neutral medium and study two different flow velocities -- a slow flow with $v_{\rm flow} = 6.8$~km s$^{-1}$ and a fast flow with $v_{\rm flow} = 13.6$~km s$^{-1}$. The clouds formed by the collision of these flows form stars, with star formation beginning after 16~Myr in the case of the slower flow, but after only 4.4~Myr in the case of the faster flow. In both flows, the formation of CO-dominated regions occurs only around 2 Myr before the onset of star formation. Prior to this, the clouds produce very little emission in the $J = 1 \rightarrow 0$ transition line of CO, and would probably not be identified as molecular clouds in observational surveys.  In contrast, our models show that H$_{2}$-dominated regions can form much earlier, with the timing depending on the details of the flow. In the case of the slow flow, small pockets of gas become fully molecular around 10~Myr before star formation begins, while in the fast flow, the first H$_{2}$-dominated regions occur around 3~Myr before the first prestellar cores form. Our results are consistent with models of molecular cloud formation in which the clouds are dominated by ``dark'' molecular gas for a considerable proportion of their assembly history.
\end{abstract}


\section{Introduction}
\label{sec:intro}

The life-cycle of molecular clouds sets the time-frame over which the star formation process can occur. As such, molecular cloud formation has received much attention in recent years, as we try to distinguish between theories for `rapid' or `slow' star formation. In a simplified picture, the cycle can be thought of as comprising three distinct phases: assembly of the cloud, the formation of stars within the cloud, and the cloud's eventual dispersal (often linked to the termination of the star formation process). While we have made some progress on understanding the latter two stages of the cloud life-cycle \citep[see e.g.][]{tamburro08, Jeffries2011}, the 
first stage is more uncertain, and has prompted much debate 
\citep[see e.g.][]{bp99,elmegreen00,hartmann01,tassis04,mtk06,blitz07,elmegreen07,tamburro08,pagani11}.

The idea of the colliding flow model for the formation of giant molecular clouds (GMCs) offers a simplified picture of how clouds can be assembled: two warm, marginally supersonic flows of gas collide head-on, and a dense molecular cloud builds up in the shocked layer \citep{elmegreen93,WalderFolini1998a, WalderFolini1998b,hp99,hp00,ki00,ki02,berg04,ah05,enrique2006,Heitsch2006,Hennebelle2008, banerjee09}. This model has several attractive features. First, there is no need to invoke some {\it ad hoc} model for driving turbulence in the cloud, as the collision between the flows and the thermal instability that occurs within the shocked gas naturally yield a linewidth-size relationship comparable to those observed within Galactic GMCs \citep{ki02,bonnell06, db07, Hennebelle2007, hhb2008}. Second, to some zeroth approximation, the model probably captures how most molecular clouds form; or at least those in spiral galaxies, where the gas streams have ample opportunity to collide \citep{dobbs2008}. Furthermore, it seems that the timescale involved in the buildup of dense regions is reasonably insensitive to the strength of the magnetic field \citep{banerjee09}. Finally, the timescale for cloud formation, and the onset of star formation can then be identified with the timescale for the onset of the gravitational instability in the shocked layer. The model is very much in the spirit of `fast' or `dynamical' star formation, where the whole process of cloud assembly and gravitational collapse to form stars occurs on the dynamical timescale -- or `crossing time' -- of the system \citep{bp99,elmegreen00,hartmann01,vs07}, yielding groups of  stars that have age-spreads smaller than their crossing times (see also the discussions in \citealt{mlk2004} and \citealt{mo2007}).

A crucial feature of the colliding flow model is the better-than-isothermal cooling that is offered by the interstellar medium as the gas is compressed. In particular, the cooling provided by C$^+$ via its 158~$\mu$m fine-structure transition permits the gas to undergo an isobaric thermal instability \citep{field65} that can produce a `two-phase' interstellar medium (ISM), with warm diffuse gas in pressure equilibrium with cold dense gas. In the conditions appropriate for the local ISM, the warm phase, 
known as the warm neutral medium (WNM), has a temperature of around 5000\,K  and a number density of around 1~cm$^{-3}$, while the cold phase (the cold neutral medium, or CNM) has a temperature of around 50~K and a number density of around 100~cm$^{-3}$ \citep{wolf95,wolf03,fer01}.

% *** This seems largely to be restating paragraph 2, and therefore winds up feeling somewhat redundant. If there's anything
% here you feel we really need to keep, then perhaps we should make it part of paragraph 2 ***
%
%The colliding-flow picture of cloud formation has had a number of successes, including the driving of turbulence at small scales (REFS), and the formation of dense, cold pockets of gas that have similar properties to observed prestellar cores (MORE REFS). Also, it appears that magnetic field do not prevent the formation of the clouds, with even marginally subcritical flows forming stars when ambipolar diffusion is taken into account. Perhaps more fundamentally however, the flows provide a simple mechanism to form groups of stars that have age-spreads smaller than their crossing times (Cite Heitsch), and thus solving the timescale problem identified by Ballesteros-Paredes et al 1999 and \citet{hartmann01}.  Considering the results of these studies, it seems that the  picture of `rapid' molecular cloud formation  -- and with it, the implication of `rapid' {\em star formation} -- is well supported, provided the conditions described by the colliding flow set-up can occur in nature.

An important issue that remains to be addressed is the ability of molecular gas to form in these flows. 
In order to form a GMC, it is necessary  to convert a significant mass of hydrogen from atomic to molecular form,
and to do this in a timescale that is comparable to or less than the assembly time of the GMC. However, with the exception of a few one-dimensional
studies \citep[e.g.][]{hp99,hp00,ki00,berg04}, most previous simulations of colliding flows have not 
modelled the chemical evolution of the gas. Therefore, although they have demonstrated that large clouds of cold 
gas can be built up in this fashion, they have been unable to address what fraction of this gas will be molecular and 
what fraction will remain atomic. Also, they have have been unable to make solid observational predictions for what the cloud formation process should look like.

The formation of H$_{2}$ in the turbulent ISM has been modelled in a number of simulations that do not directly
investigate the colliding flow paradigm. For example, on large scales, the work of \citet{dobbs08} has demonstrated that `molecular' clouds can be formed via compression of the gas in the spiral arms of galaxies. For scales greater than around 
10~pc, this study demonstrated that H$_2$ can form on timescales shorter than the crossing time of the spiral-arm passage. 
On smaller scales, \citet{gm07b} have shown that H$_2$ can easily form on a dynamical timescale in gas with number densities of around 100 cm$^{-3}$, thanks to the boost to the H$_{2}$ formation rate provided by the many transient
turbulent density compressions. Similar results were also recently reported by \citet{micic12} for the case of 
compressively-driven turbulent flows, and \citet{mg12} have shown that the ratios of H$_{2}$ to atomic hydrogen produced
by these small-scale models after a single gravitational free-fall time are in good agreement with the observed values \citep{br04,br06}. If one combines the results of these studies, then it would seem clear that forming clouds of H$_2$ is fairly easy, and that the colliding flow model should have no difficulty in producing large quantities of gravitationally-unstable molecular gas.

However it is via CO, not H$_2$, that molecular clouds are observationally defined, and so any formation model must necessarily produce enough CO to allow the cloud to be identified as a GMC by observational surveys. To date, the formation of CO has only been studied directly in models of turbulent, pre-assembled clouds (see e.g.\ \citealt{g10,gm11,gc12a,gc12b, Shetty2011a, Shetty2011b} for models that treat both the hydrodynamics and the chemistry, or \citealt{roellig07} for an entry point into the extensive astrochemical literature dealing with static cloud models). The
issues of how the CO formation timescale relates to the timescale on which the gas is assembled, or how it relates to the emergence of star-forming regions within the cloud have yet to be explored self-consistently. The closest that any previous
colliding-flow study has come to this is the work of  \citet[][hereafter HH08]{hh08}. They took a simplified approach to this problem, and suggested that any region in a colliding flow that has a temperature of less than 50~K and a mean visual extinction of
$A_{\rm V} = 1$ or higher will have a high CO abundance. Using this approximation, they found that colliding flows could produce large, observable molecular clouds on timescales shorter than the star formation timescale (by around 4--5~Myr 
in the flows that they looked at). However, given the non-equilibrium nature of the CO chemistry, and the complicated relationship between density, temperature and extinction found by \citet{g10}, it is unclear whether the assumptions made by HH08 are correct, and it is therefore worth revisiting this type of calculation with a more sophisticated treatment of the chemistry.

With this in mind, we revisit the colliding-flow model of cloud formation with a self-consistent time-dependent chemical network -- coupled to a thermodynamical model of the ISM -- that runs alongside the hydrodynamics. The chemical network is able to evolve the abundances of key species such as H, H$^+$, H$_2$, C, C$^+$ and CO, and allows us to follow the transition of the gas from the WNM to the CNM and onwards to the cold molecular cores that are the sites of star formation.

\begin{figure*}
\centerline{
  \includegraphics[width=6.4in]{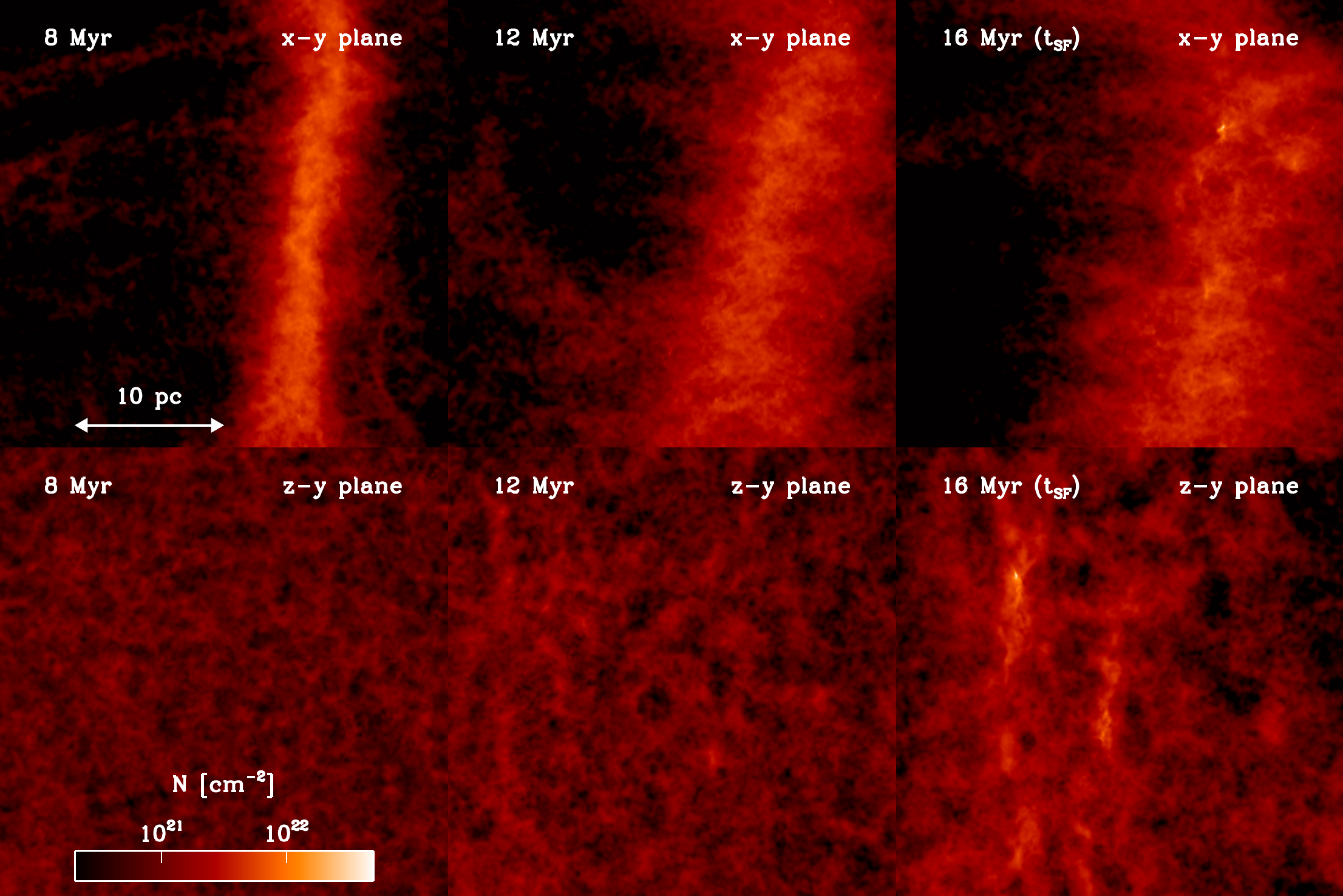}
	 }
\caption{Column density images showing the evolution of the 6.8 km s$^{-1}$ (slow) flow at 3 points before the formation of the first sink particle. The top images show the $x$-$y$ plane, perpendicular to the flow, while the bottom panels show the cloud as seen looking down the $x$-axis onto the $z$-$y$ plane -- that is, looking {\em along} the flow, onto the shock plane. }
\label{fig:slowpics}
\end{figure*}

\begin{figure*}
\centerline{
  \includegraphics[width=6.4in]{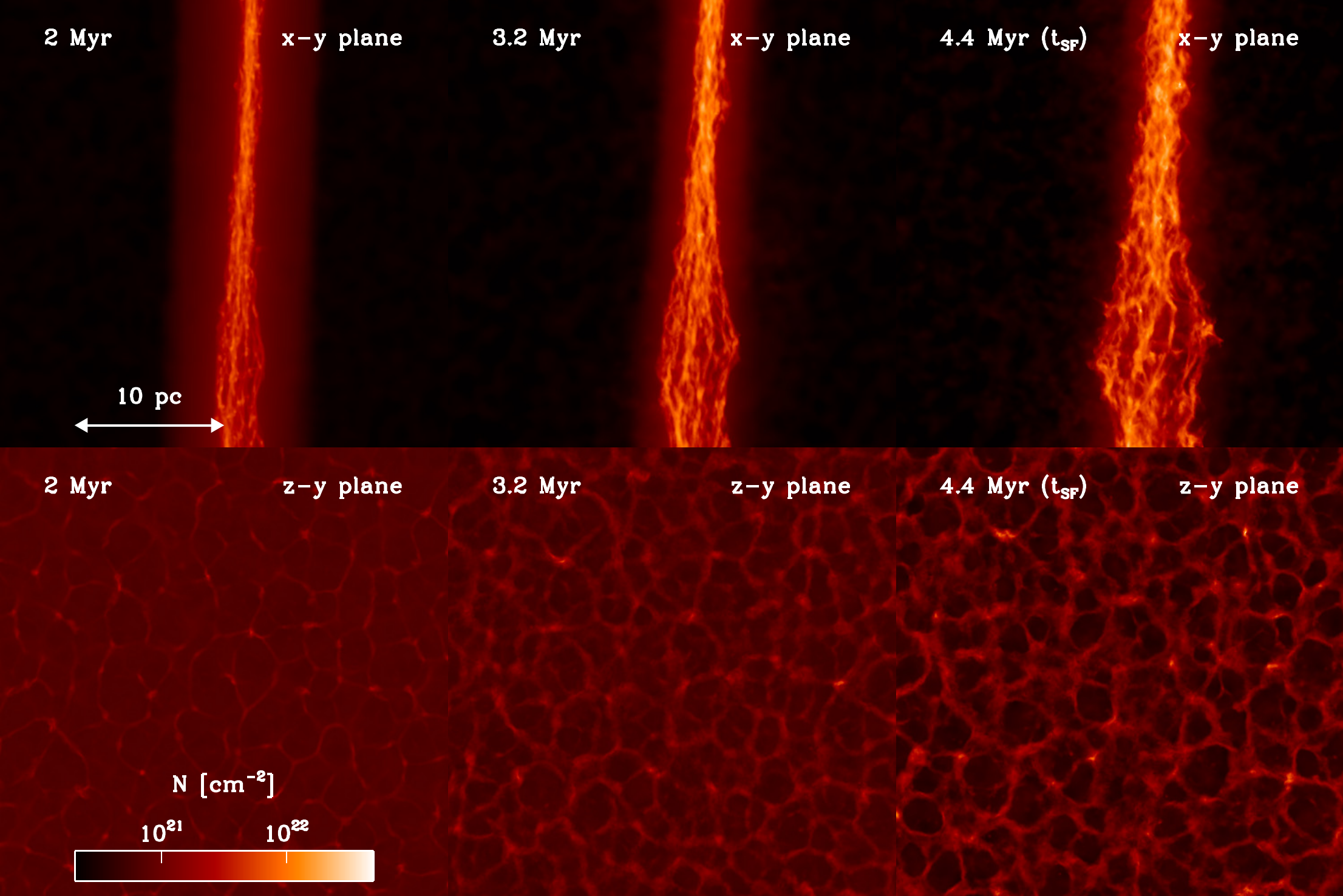}
 }
\caption{As Figure~\ref{fig:slowpics}, but for the 13.6 km s$^{-1}$ (fast) flow. Note that the output times examined here are different from those in Figure~\ref{fig:slowpics}.}
%Column density images showing the evolution of the 13.6 km s$^{-1}$ (slow) flow at 3 points before the formation of the first sink particle. The top images show the x-y plane, perpendicular to the flow, while the bottom panels show the cloud as seen looking down the x-axis onto the z-y plane -- that is, looking {\em down} the flow, {\em onto} the shock plane. }
\label{fig:fastpics}
\end{figure*}

%
%%% Section on the simulations.
%

\section{Simulations}
\label{sec:sims}

\subsection{The numerical model}
\label{sec:numerics}
We model the evolution of the gas in our simulations using a modified version of the {\sc Gadget2 } smoothed particle hydrodynamics code \citep{springel05}. We have extended the publicly available code in several respects. First, we have added a sink particle implementation, based on the prescription in \citet{bbp95}, to allow us to follow the evolution of the gas beyond the point at which the first protostar forms. Our particular implementation is the one first described in \citet{jappsen05}. In addition, we have implemented an external pressure term \citep[e.g.][]{benz90} into the SPH equations that enables us to model a constant pressure boundary, as opposed to the vacuum, periodic, or `fixed' boundary conditions that are the only choices available in the standard version of {\sc Gadget2}. The details of this term can be found in \citet{clark11}.

Our treatment of the gas-phase chemistry combines the hydrogen chemistry network introduced in \citet{gm07a,gm07b} with the treatment of CO formation and destruction proposed by \citet{nl99}. This combined network was tested against other simplified chemical networks for CO formation and destruction by \citet{gc12b}, who showed that it does a very good job of reproducing the results of more complex networks (e.g.\ \citealt{g10}) for C$^{+}$, C and CO, while incurring only one-third of the computational cost. Further details regarding our treatment of the chemistry can be found in \citet{gc12b}. Note that the simulations presented in this paper do not include the effects of the freeze-out of CO onto dust grains. In Galactic star-forming clouds, CO freeze-out has been shown to have only a very minor effect on the thermal balance of the gas \citep{gold01}.

To model how the interstellar radiation field (ISRF) penetrates into the cloud, we make use of the \textsc{TreeCol} algorithm \citep{cgk12}. This algorithm provides each SPH particle with a $4\pi$ steradian column density map of the sky, stored in 48 \textsc{Healpix} pixels \citep{healpix}. The column density information can then be used to calculate the attenuated ISRF at the location of each particle -- information that is vital for calculating the dust temperatures and photoelectric emission rates, as well as computing the photo-dissociation rates in the chemical networks. To convert from the column density to the dust extinction, we adopt the relationship $A_{\rm V} = 5.348 \times 10^{-22} [N_{\rm H, tot} / 1 \: {\rm cm^{-2}}]$ \citep{bsd78,db96},
where $N_{\rm H, tot}$ is the column density of hydrogen nuclei. We also employ \textsc{TreeCol} to compute maps of the H$_{2}$ and CO column densities, and use these to determine shielding factors due to H$_{2}$ self-shielding, CO self-shielding and the shielding of CO by H$_{2}$. For H$_{2}$ self-shielding, we take the a shielding function given in \citet{db96}, while for the other two processes, we make use of the values tabulated by \citet{lee96}. We would not expect to 
obtain significantly different results were we to use the updated treatment of CO self-shielding presented in \citet{visser09}.

\subsection{Details of the colliding flow}
\label{sec:flowic}

We start our flow at a number density of 1 cm$^{-3}$ (note that we will refer to `number density' simply as `density' in what follows) and a temperature of 5000\,K. The geometry of the flow is a rectangular cuboid of length $\pm$ 111.5 pc in the $x$-direction and a width of $\pm$ 36 pc in $y$ and $z$. The total mass in this volume is $4 \times 10^4$ \solmasp. We apply a flow of $+v_{\rm flow}$ km s$^{-1}$ for particles at negative $x$ and $-v_{\rm flow}$ for particles at positive $x$, such that the flows move along the $x$-axis towards zero.  In this paper we look at two flow speeds,  $6.8$ km s$^{-1}$, and $13.6$ km s$^{-1}$, which we will refer to as the `slow' and `fast' flows respectively. The time taken for all of the gas to reach the centre is therefore $\sim 16.6$ Myr in the slow case, and $\sim 8.3$ Myr in the fast case. The isothermal sound speed in the gas is initially 5.6 km s$^{-1}$, so our flow velocities have initial Mach numbers of $\mathcal{M} = 1.22$ and $\mathcal{M} = 2.62$, for the slow and fast flows, respectively. We also impose a low level of turbulence on the flows that has a power spectum of $P(k) \propto k^{-4}$, with turbulent velocities scaled such that the rms turbulent velocity is given by $v_{\rm rms} = 0.2\,c_{\rm s} $. 

We use 8,000,000 SPH particles to model the gas, and the particles are initially placed on a grid. When computing fluid properties, we require that each particle has 70 neighbours. The resulting mass resolution is therefore 0.7 \solmas \citep{bb97}. We have found in previous work that a mass
resolution of around this value is sufficient for determining the star formation rate \citep{gc12a}, but note that it is probably not sufficient to allow us to draw any strong conclusions regarding the stellar initial mass function (IMF). We distribute the particles on a uniform grid at the start of the simulation. We then evolve the flow, using a value of 1.5 for the {\sc Gadget2} viscosity parameter. We prevent immediate expansion of the flow through use of a confining external pressure, which we set to be $P_{\rm ext} = 5000\,k_{\rm B}$, where $k_{\rm B}$ is the Boltzmann constant. This value is comparable to the expected equilibrium pressure in the warm neutral gas given our choices for the metallicity, UV field strength and cosmic ray ionization rate described below.

We consider solar metallicity gas, and adopt values for the initial elemental abundances of carbon, silicon and oxygen given
by \citet{sem00}: $x_{\rm C, tot} = 1.4 \times 10^{-4}$, $x_{\rm Si, tot} = 1.5 \times 10^{-5}$ and $x_{\rm O, tot} = 3.2 \times 10^{-4}$, where $x_{i}$ denotes a fractional abundance, by number, relative to the number of hydrogen nuclei. We assume that the carbon and silicon start in the form of C$^{+}$ and Si$^{+}$, owing to photoionization by the ISRF, and that the oxygen starts in neutral atomic form. The initial H$^{+}$ abundance in the gas was set to $x_{\rm H^{+}} = 6.5 \times 10^{-3}$, which is approximately the value yielded by equilibrium between cosmic ray ionization and radiative recombination. The remainder of the hydrogen was assumed to be in atomic form. The dust-to-gas ratio was taken to be 0.01, the canonical value for the local ISM. For the dust opacities, we use two main sources of data. For the long wavelength opacities ($\lambda > 1 \: \mu{\rm m}$), we use the values given by \citet{oh94} for non-coagulated grains with thick ice mantles. For shorter wavelengths, we use the values quoted in \citet{mmp83}.  In reality, dust grains in the warm ISM are highly unlikely to have ice mantles, but we note that in this regime, the particular values chosen for the long wavelength grain opacities have very little influence on the thermal  evolution of the gas. They become significant only in regions with high dust extinctions, where both gas and dust are typically very cold, and in these circumstances the grains will likely develop ice mantles. 

For the interstellar radiation field, we use the parameterisation given by \citet{bl94} for optical and longer wavelengths, and 
the \citet{dr78} fit in the ultraviolet regime. The cosmic ray ionization rate of atomic hydrogen in all of our simulated clouds was chosen to be $\zeta_{\rm H} = 3 \times 10^{-17} \: {\rm s^{-1}}$. The cosmic ray ionisation rates for the other major chemical species tracked in our chemical model were assumed to have the same ratio relative to the rate for atomic hydrogen as given in the UMIST99 chemical database \citep{teu00}.

As we include the effects of self-gravity in our simulations, the gas is eventually able to form bound regions that start to undergo run-away gravitational collapse. In order to allow us to continue the simulations beyond the point at which the first of these regions collapses, we make use of sink particles. Regions that reach a density of $10^6$ cm$^{-3}$ are considered as candidates for sink particle creation. To ensure that sink particles are not created out of transiently collapsing gas that should later re-expand, the SPH particles at the centre of the region have to pass a series of tests to verify that they are actually gravitationally bound and collapsing, as outlined by \citet{bbp95}. In our study, the sink particles have an accretion radius of 600 AU, and their gravitational interactions are softened on a scale of 100 AU. We also limit the distance between a new sink forming and a neighbouring sink to 1200 AU. Sink particles are permitted to accrete SPH particles once the gas falls inside the accretion radius, provided that the candidate SPH particle is not only bound to the candidate sink particle, but more bound to that sink than to any other. The SPH particle's mass and momentum are then transferred to the sink particle. In this manner, the mass in sink particles can be used as a proxy for the mass in stars. 

In addition to the two simulations discussed in the following sections, we also performed two additional simulations that we do not  discuss in detail here. The first of these was a lower resolution simulation with only 200,000 SPH particles that we used to check our numerical convergence. We found that while many of the features that we discuss in this paper were delayed with respect to the higher resolution runs, the basic physical and chemical traits were the same. The delay -- which was around 1 million years -- is expected, since when the resolution is poor, the small scale fluctuations are washed out, and the shock heating is spread over a larger mass of gas. The second additional simulation employed a different realisation of the random turbulent field that we use to inject inhomogeneities into the flow. Again, we did not find any significant differences that would alter the conclusions that we draw below.

\section{Evolution of the flows}
%Density and temperature evolution}
\label{sec:temprho}

\begin{figure}
\includegraphics[width=3.3in]{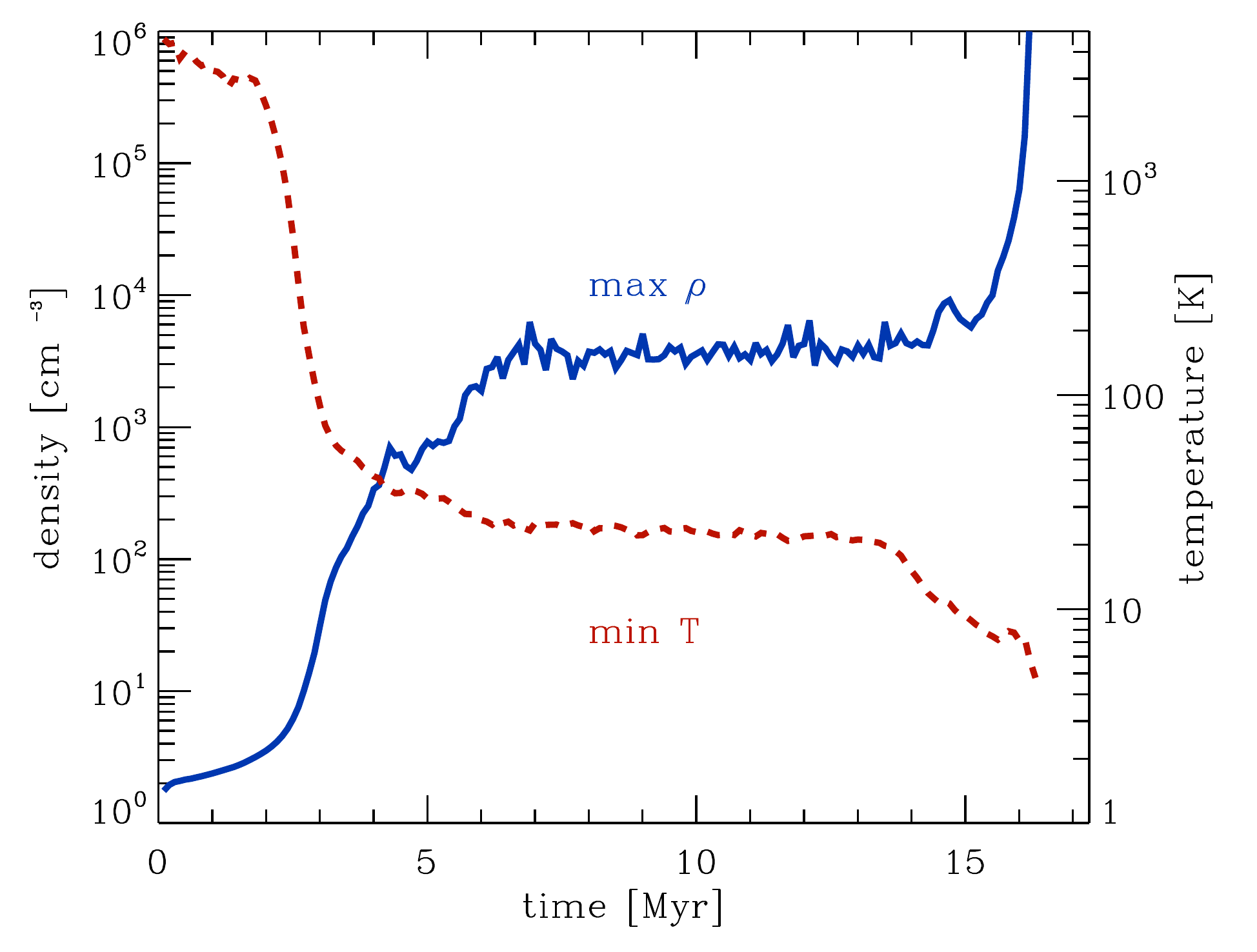}
\includegraphics[width=3.3in]{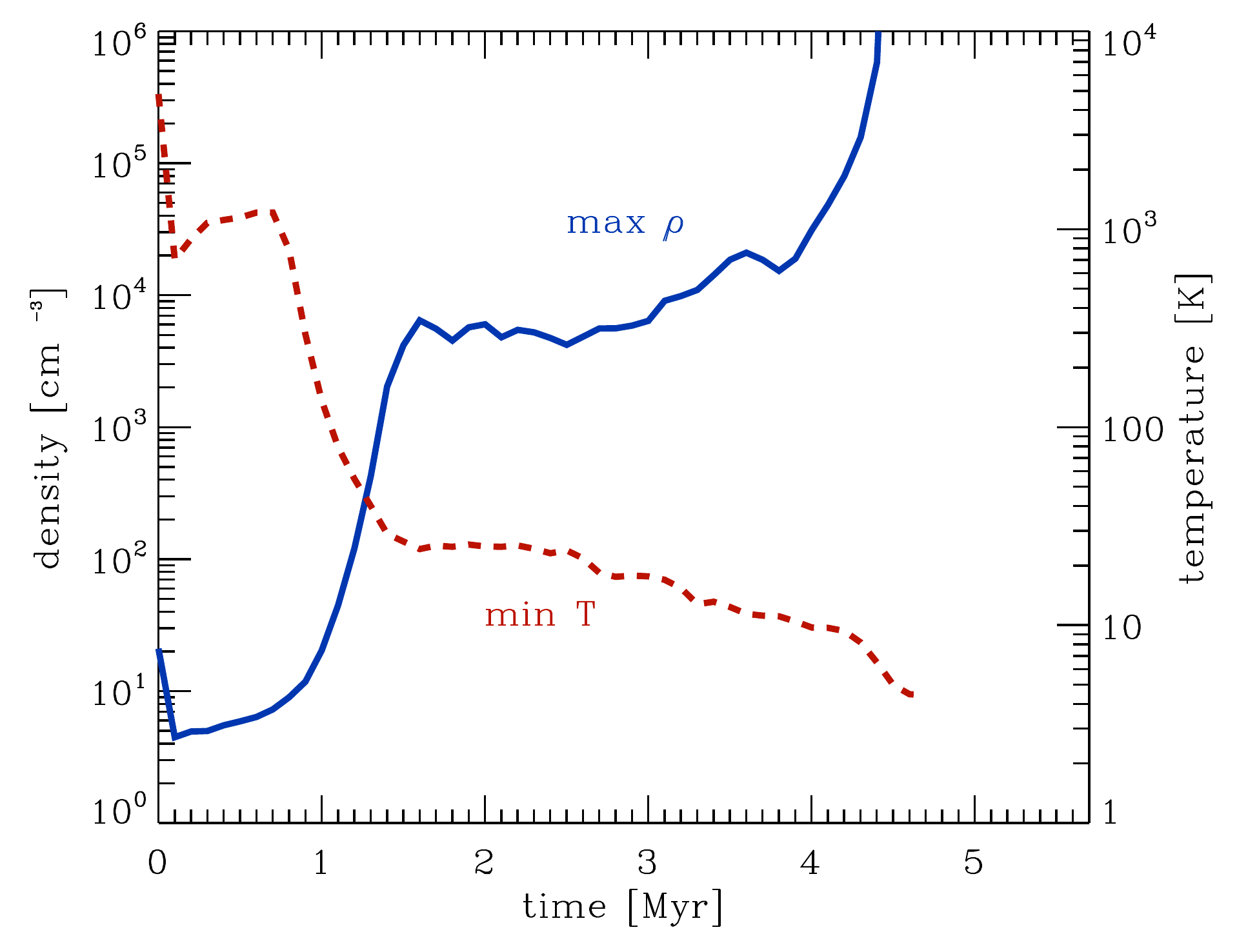}
\caption{Evolution with time of the maximum density (blue, solid line) and minimum temperature (red, dashed line) in the slow flow (top panel) and the fast flow (bottom panel). Note that at any given instant, the coldest SPH particle is not necessarily the densest, and so the lines plotted are strictly independent of one another.}
\label{fig:rhotminmax} 
\end{figure}

\begin{figure*}
\centerline{
  \includegraphics[width=2.8in]{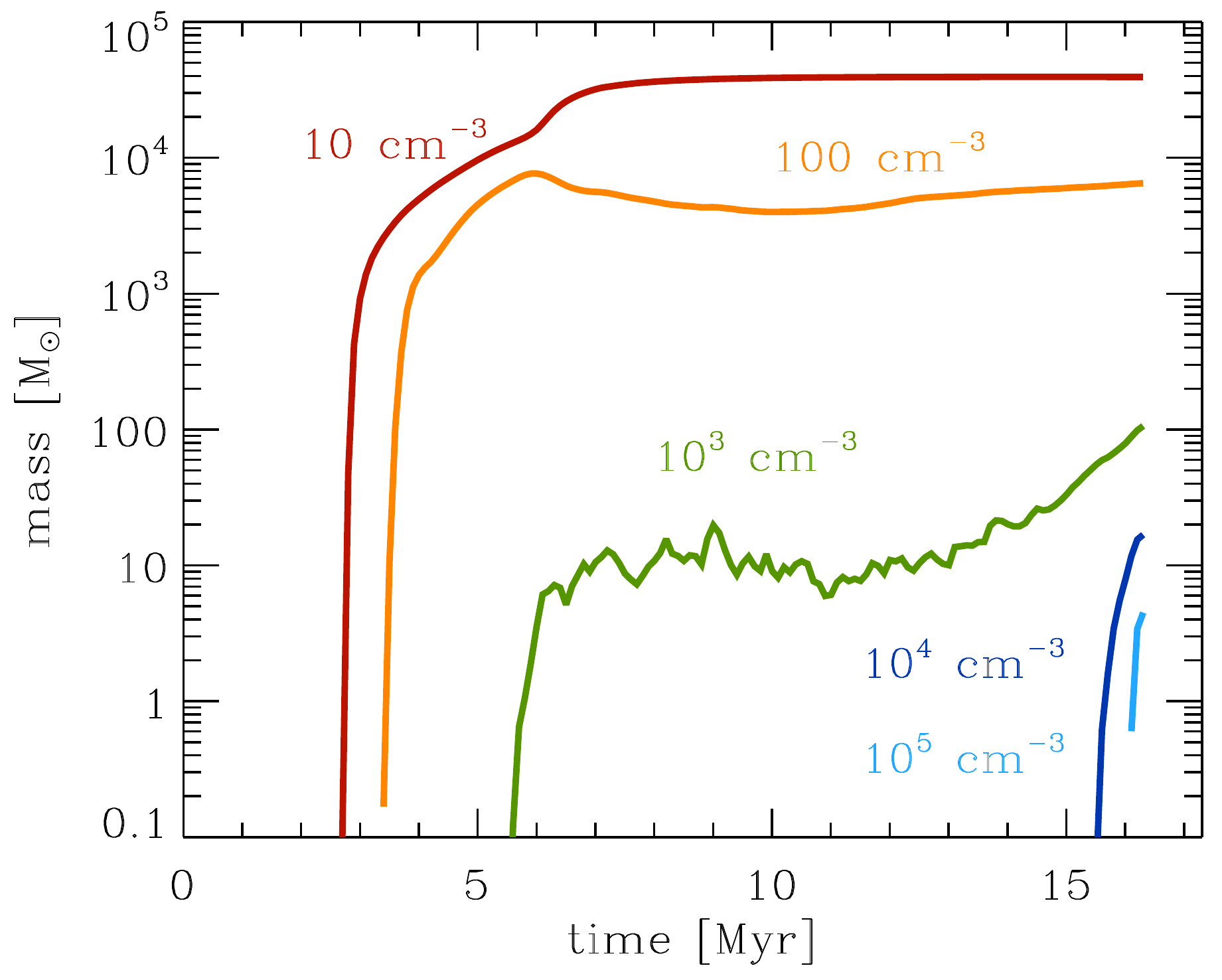}
  \includegraphics[width=2.8in]{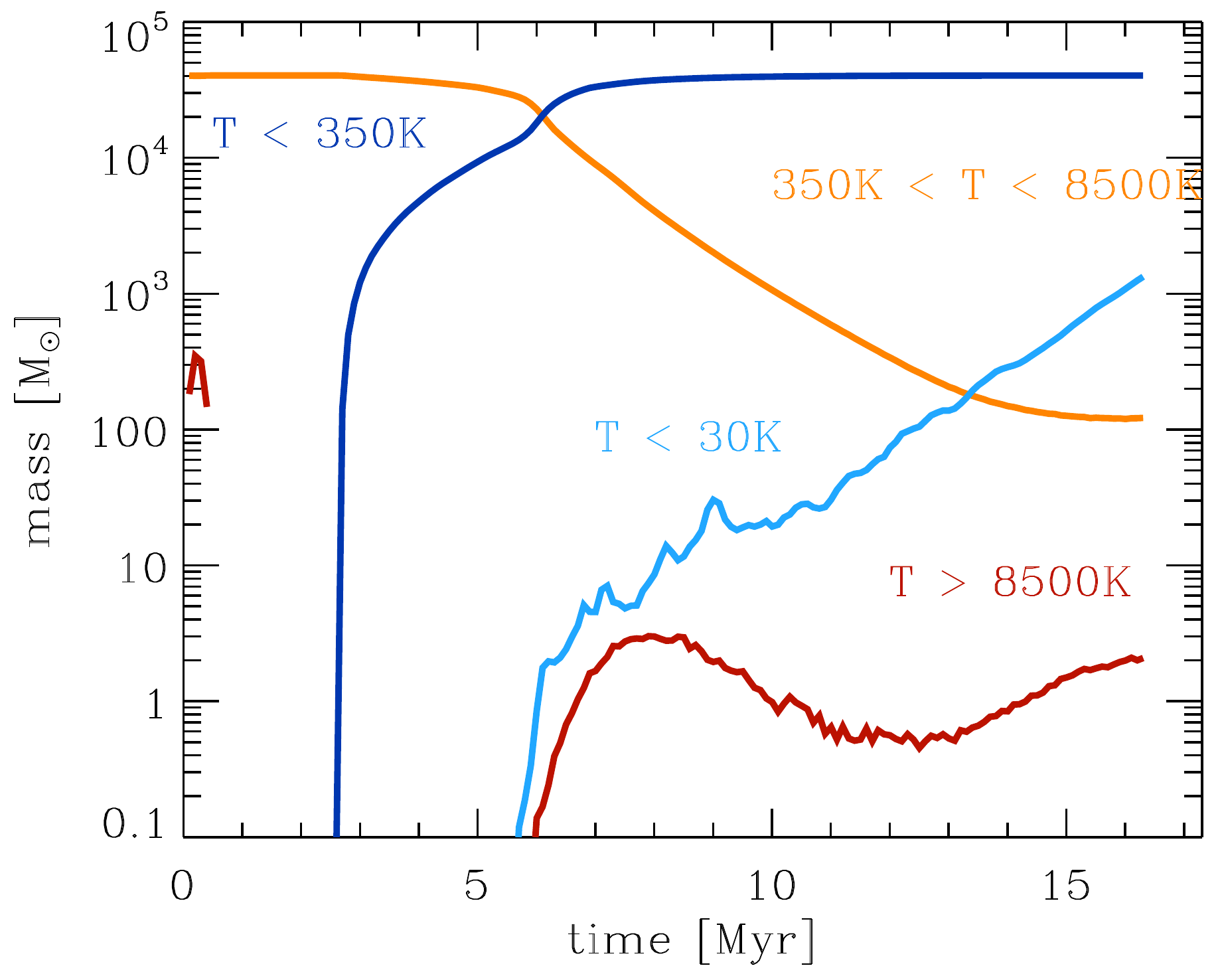}
 }
 \centerline{
  \includegraphics[width=2.8in]{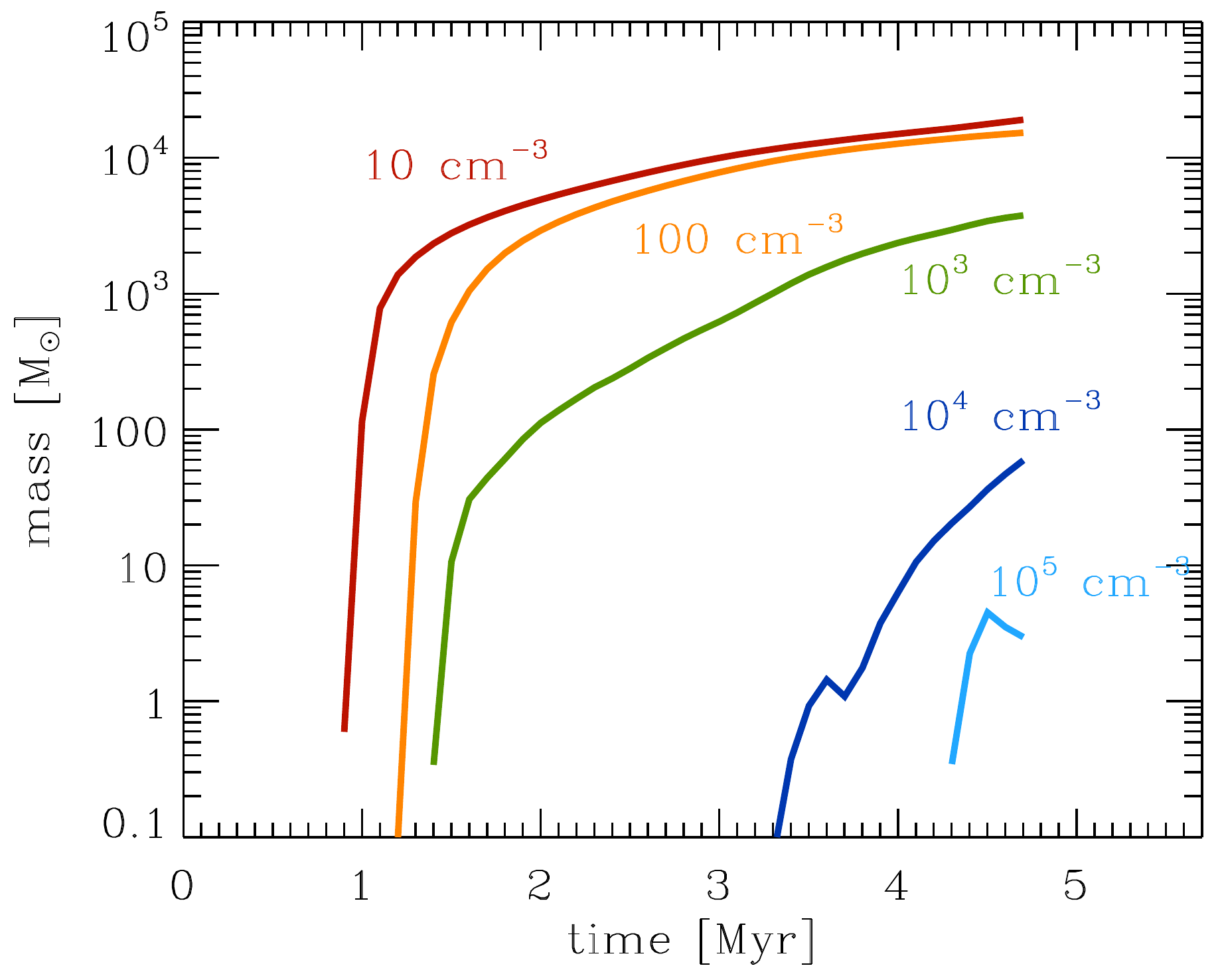}
  \includegraphics[width=2.8in]{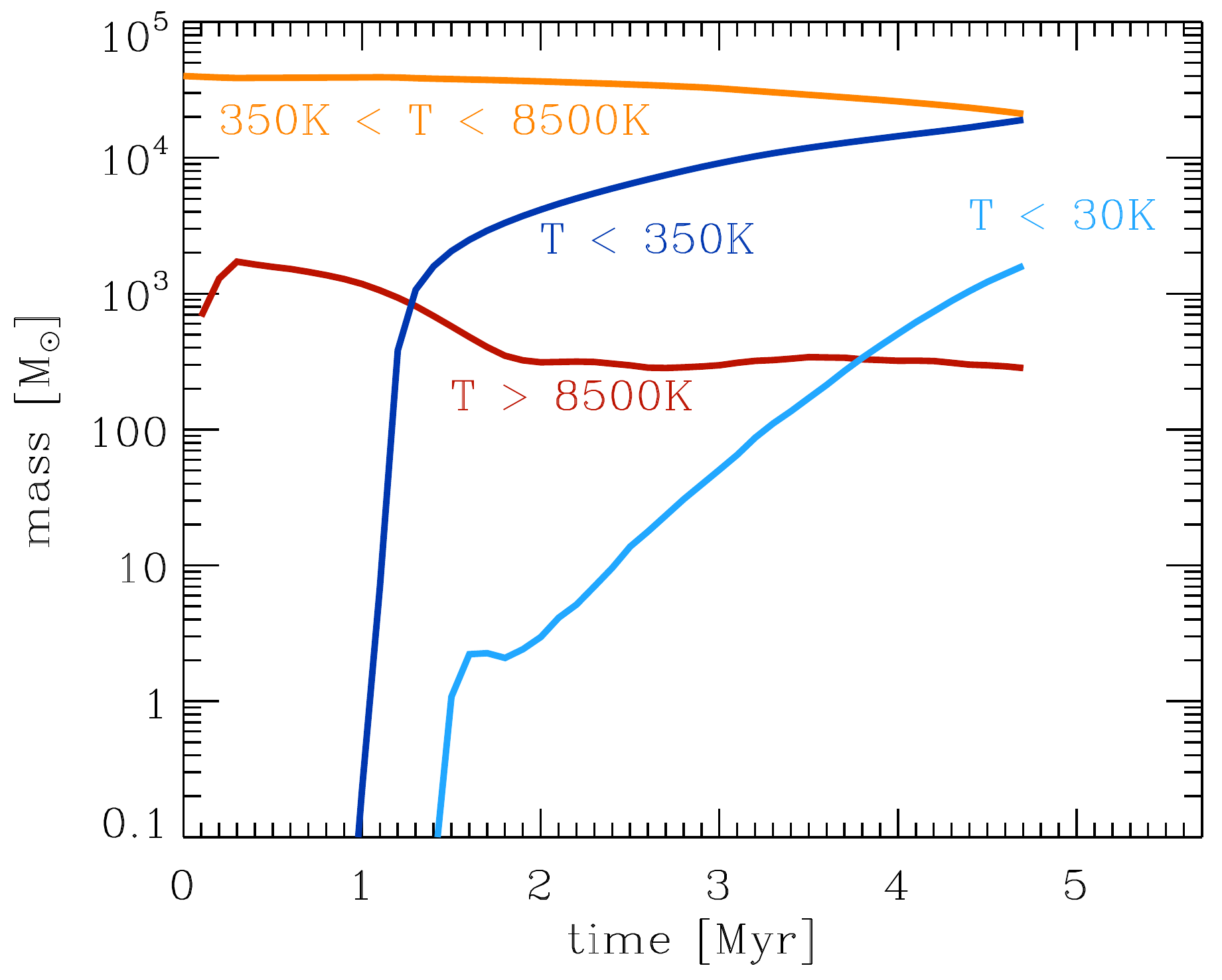}
 }
\caption{The plots on the left show the mass of gas that sits above a given density threshold as function of time, while those on the right show the fractions of the cloud mass that lie within given temperature ranges. The top panels are for the 6.8 km s$^{-1}$ flow, and the bottom panels follow the 13.6 km s $^{-1}$ flow. Note that the scales on the time axis differ between the top and the bottom panels.}
\label{fig:rhotemp} 
\end{figure*}

In Figures~\ref{fig:slowpics} and \ref{fig:fastpics}, we show images of the column density of the gas, looking both 
perpendicular to the axis of the flow (upper panels) and along the axis (lower panels).
Images are shown for three output times: the time at
which the first sink particle forms (hereafter referred to as $t_{\rm SF}$, denoting the onset of star formation in the clouds),
plus two earlier times that are approximately 50\% and 75\% of $t_{\rm SF}$. The images show a sequence of events similar
to those seen in previous colliding flow models. In the region where the flows collide, a thermal instability is triggered which
produces a dense, cold layer of gas with significant turbulent sub-structure. Initially, the small-scale structure in this cold layer is not self-gravitating, but eventually some small dense regions become gravitationally bound and go into run-away gravitational collapse, leading to the formation of stars. In the case of the slow flow, star formation begins after roughly 16~Myr, while the fast flow begins to form stars after only 4.4~Myr. 

A more quantitative overview of the evolution of the flow is provided by an examination of the density and temperature
evolution of the gas. As molecular clouds are typically much denser and colder than our initial conditions, we first look at the evolution of the maximum density and minimum temperature in the gas, and how this develops as the flows come together. These quantities are plotted in Figure \ref{fig:rhotminmax}, and were calculated for the densest and coldest SPH particles in the simulation at any point. Note that although dense gas tends to be cold, the densest particle is not necessarily the coldest particle, and so at any given time, the lines can be defined by different SPH particles. 

Initially, the two flows exhibit very similar behaviour: the minimum gas temperature drops to around 30 K within a few Myr, accompanied by a gradual rise in the peak density. The time taken for this initial phase of cooling is longer for the slow 
flow than for the fast flow. This can be understood as a consequence of the difference in the initial post-shock density in the
two cases. The shocked gas in the collision region of the fast flow has a higher density, owing to the higher Mach number
of the flow, and hence cools more rapidly than is the case with the slow flow. 

At first, the shocked gas has a higher thermal pressure than the unshocked gas surrounding it, although it is prevented from
expanding in the $x$-direction by the ram pressure of the inflowing gas. As the gas cools, however, it first re-establishes thermal pressure equilibrium with its surroundings and then maintains this equilibrium by increasing its density as its temperature decreases. By the time the gas has cooled to 30~K, its mean density has increased by roughly two orders of magnitude (see Figure~\ref{fig:rhotemp}). The cold gas has also become turbulent, leading to a much larger increase in the peak density of
the gas, owing to the influence of turbulent compressions occurring within the cold slab. 

The initial cooling phase is complete after around 4~Myr for the slow flow, and roughly 1.5~Myr for the fast flow. At this 
point, there is a marked change in the behaviour of the maximum density and minimum temperature. In the slow flow, we
see that both quantities remain roughly constant for around 10~Myr. In the fast flow, a similar phase occurs for roughly 
2.5~Myr, although in this case there is a slow but steady decrease in the minimum temperature and a similar increase
in the maximum density. We can identify this phase in the evolution of the cloud as one in which turbulence is more
important than self-gravity for determining the peak density. Although dense substructure exists within the cloud, 
created by the combination of thermal instability and turbulent compressions, the majority of this substructure is not
gravitationally bound (see, for example, the discussion in \citealt{banerjee09}).

\begin{figure}
\centerline{
  \includegraphics[width=3.2in]{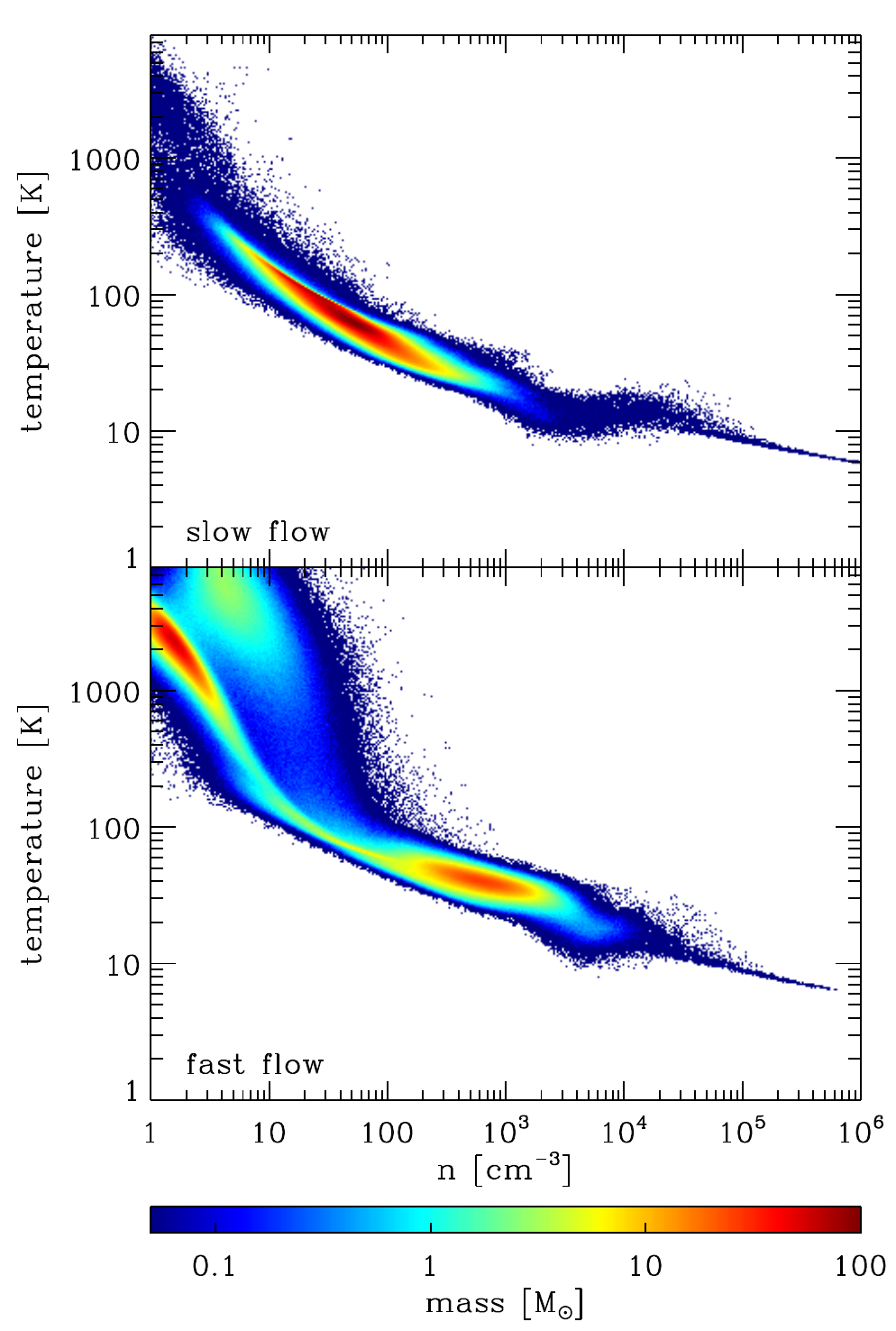}
}
\caption{The gas temperature-density distribution in the flows at the onset of star formation.}
\label{fig:rhotphase} 
\end{figure}

This phase in the cloud's evolution comes to an end once gravitationally bound cores form and begin to 
undergo runaway gravitational collapse. This occurs at $t \sim 4 \: {\rm Myr}$ for the fast flow and $t \sim 16 \: {\rm Myr}$
for the slow flow, and is followed very quickly by the onset of star formation within the clouds. At the point at which the
first gravitationally-bound core goes into  runaway collapse, the minimum gas temperature has dropped to around
10~K, typical of the temperatures observed in nearby, star-forming cores \citep[see e.g.][]{bt07}. Comparing the densities 
and temperatures in Figure \ref{fig:rhotminmax} to those in Figure \ref{fig:rhotphase}, we see that the low temperatures are indeed found within the dense, collapsing cores. The lack of scatter in the density-temperature relation above 10$^4$ cm$^{-3}$ reflects the fact that the thermal coupling between the gas and the dust becomes effective at these densities within prestellar cores.

%From the discussion so far, it would seem that flows are characterised by dense cold gas, and thus we would expect star formation to occur much faster that it actually does; indeed, the gas in Figure \ref{fig:rhotminmax} appears to evolve on timescales much longer than one would expect from its free-fall time. The reason for the delay in the star formation is that there is actually very little gas in such a state, as the overwhelming majority of the gas either significantly hotter, has a lower density, or both. As in other studies, the eventual collapse of the dense cores is brought on by the collapse of a much larger, and lower density region. 

A more detailed picture of the distribution of densities and temperatures in the flow is given in Figure~\ref{fig:rhotemp}.
In the left-hand panels, we show how the mass of gas denser than a chosen density threshold $n_{\rm thr}$ evolves as
a function of time for values of $n_{\rm thr}$ ranging from $10 \: {\rm cm^{-3}}$ to $10^{5} \: {\rm cm^{-3}}$ for the two
flows. In the right-hand panels of Figure~\ref{fig:rhotemp}, we look at the temperature distribution of the gas. 
For a static gas distribution in pressure equilibrium with solar metallicity and with our chosen environmental parameters (radiation field strength, cosmic ray ionization rate, etc.), two stable thermal phases
exist: a warm phase with temperature $T > 8500 \: {\rm K}$ and a cold phase with $T < 350 \: {\rm K}$. Gas in the temperature
range $350 < T < 8500 \: {\rm K}$ is not thermally stable, and will either heat up until it joins the warm phase, or cool down until it joins the cold phase, depending on its starting density. In Figure~\ref{fig:rhotemp}, we show how the gas mass in these three regimes evolves with time in the two flows. We also track the mass of very cold gas ($T < 30 \: {\rm K}$). Most of the CO emission from local molecular clouds comes from gas with temperatures below around 20--30~K, and this very cold phase is also the most likely site of star formation in the cloud \citep[see e.g.][]{klm11}.

%Figure~\ref{fig:rhotemp} demonstrates that within 5~Myr, roughly a quarter of the total mass in the flow is in the cold phase, with density greater than 100 cm$^{-3}$ and a temperature less than 350~K. The reason why the flow speed does not strongly affect the fraction of the gas mass in this state is that the flows themselves are thermally unstable for our initial set-up. The transition to a cold, dense state occurs throughout the flow, rather than just in the shocked region, and is triggered by the subsonic turbulence present in the initial flow.

Despite this similarity, many other features of the density and temperature are seen to be very different when one compares the two flows. For example, we see that the fast flow maintains a much higher fraction of hot gas than the slow flow, reflecting the stronger shock-heating from the collision of the streams. Further, we see that the fast flow generates much more very cold ($T < 30$ K) gas, and that this cold gas makes its first appearance much earlier in the simulation. By the onset of star formation, the fast flow has over 8000 \solmas of gas in the very cold state, while the slow flow has only around 3000 \solmasp~in the same state. This means that for the fast flow, roughly 50\% of all of the gas colder than 350~K is in this
very cold state, whereas the corresponding figure for the slow flow is only 10\%. On the other hand, the fast flow also has
much more gas still in the unstable regime ($350 < T < 8500$~K), owing to the fact that in this case, star formation begins before all of the low density, unshocked gas has had time to cool sufficiently to reach the cold phase.

As well as forming significantly more cold gas, we also see that the fast flow forms more dense gas. Indeed, gas at number densities $n \geq 10^{4} \: {\rm cm^{-3}}$ -- the minimum density characteristic of the structures identified observationally as pre-stellar cores -- appears after only about 2.5 Myrs in the fast flow, in comparison to 19 Myr in the slow flow. We also see that the evolution of the density fractions in the slow flow is more stochastic, indicating that much of the structure that is formed during the collision between the streams is transient. This transient structure is either ripped apart by the ram pressure of the flows or pushed apart by internal thermal pressure and turbulent motions. In the fast flow, we do not see this behaviour;
typically, the mass above each of our threshold densities is continually rising.

%
%%% Section describing the general conditions in the clouds
%

\section{Chemical and observational timescales}
%\label{sec:chem}

In this section, we first give an overview of the general chemical evolution of the flows, and how long it takes to form a `molecular' cloud in each case -- that is, one that would be seen by an observer via CO emission. We then go on to look at how the post-processed CO maps of \citet{hh08} compare to our fully self-consistent and time-dependent treatment of the cloud chemistry. Finally, we look at how the observable properties of the CO vary with time as the clouds (and star-forming regions) are assembled.

\subsection{General chemical evolution of the flows}
\label{sec:chem}

\begin{figure*}
\centerline{
  \includegraphics[width=3.0in]{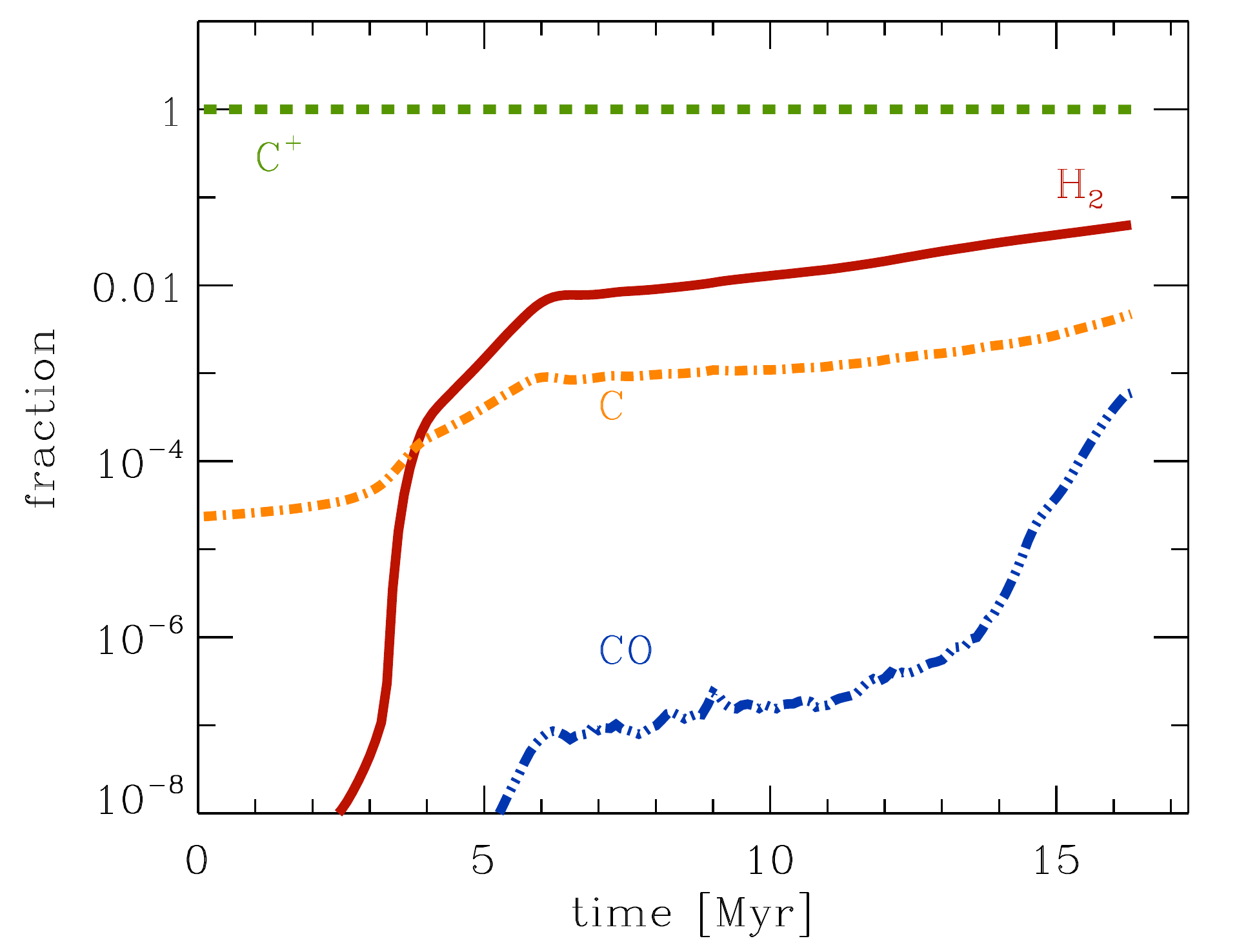}
  \includegraphics[width=3.0in]{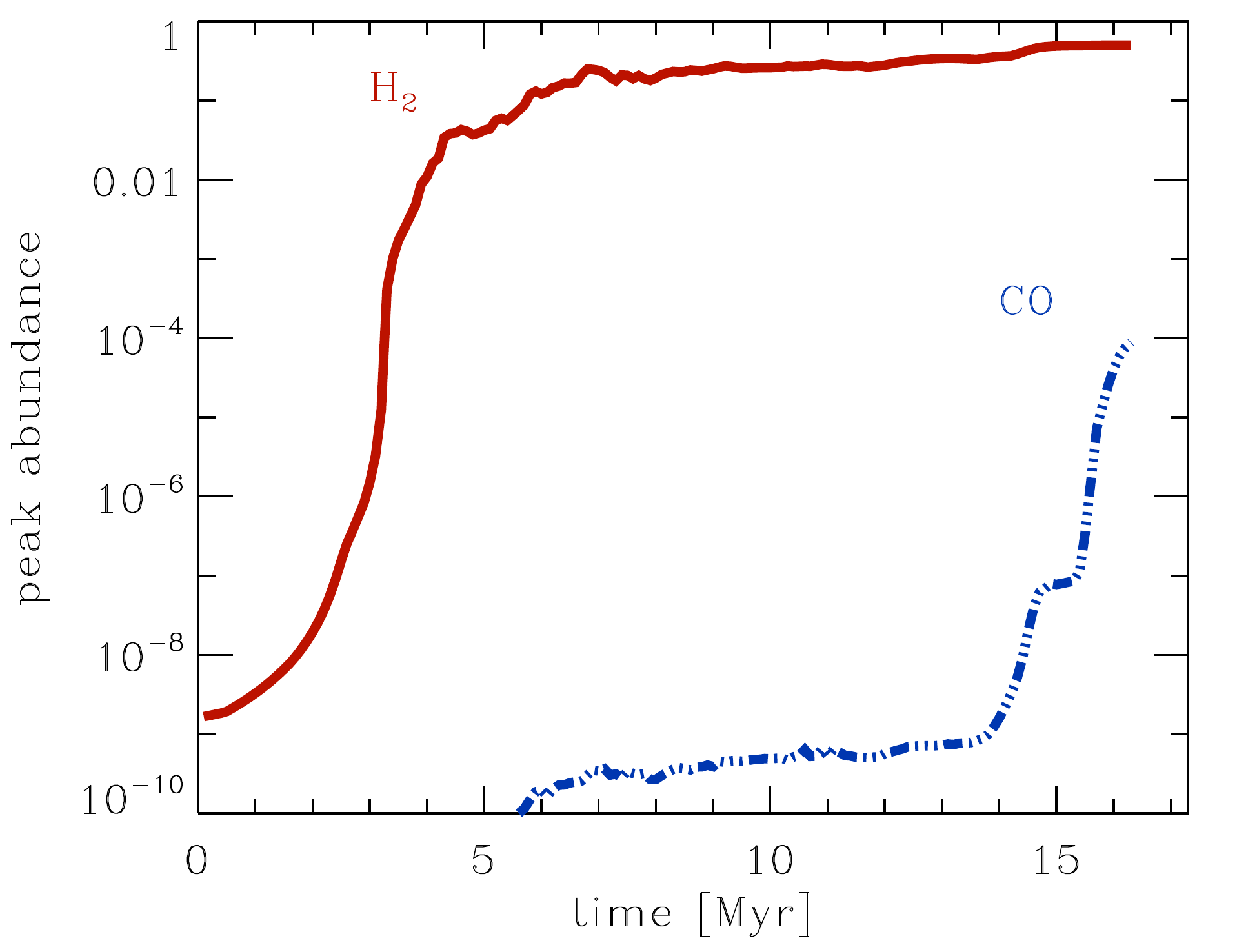}
}
\centerline{
  \includegraphics[width=3.0in]{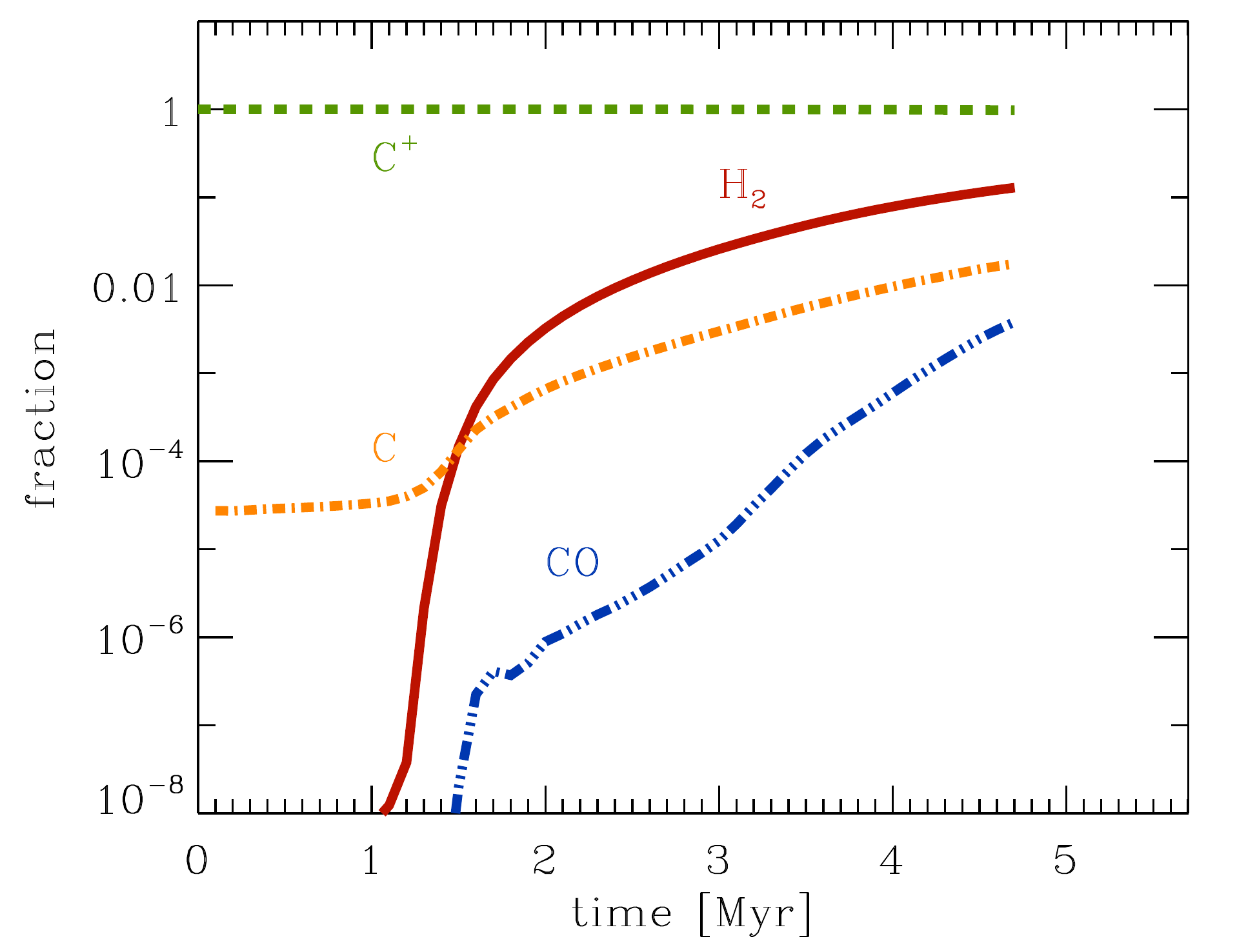}
  \includegraphics[width=3.0in]{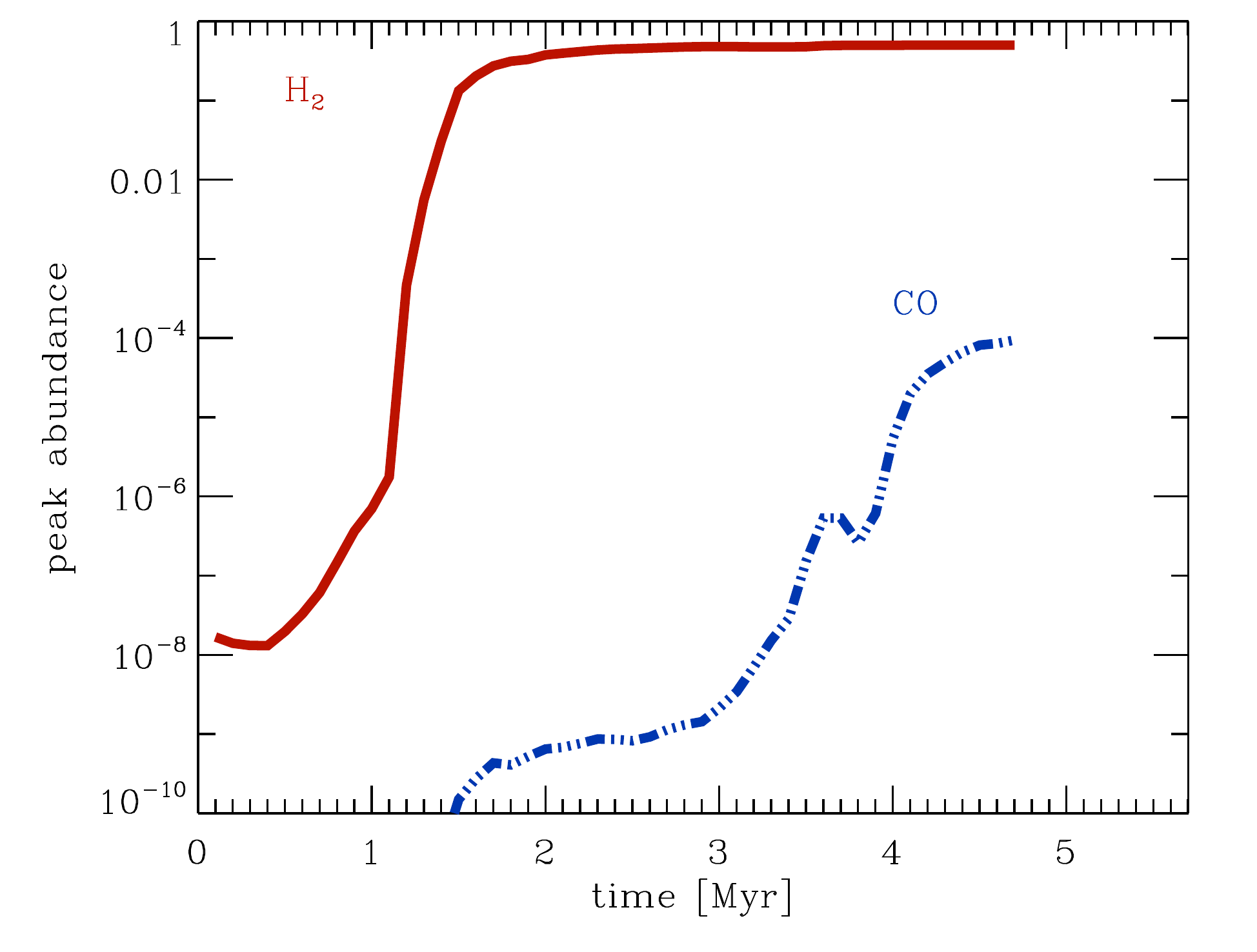}
}
\caption{Chemical evolution of the gas in the flow. In the left-hand column,  we show the time evolution of the
fraction of the total mass of  hydrogen that is in the form of H$_{2}$ (red solid line) for the 6.8 km s$^{-1}$ flow  (upper panel) 
and the $13.6 \: {\rm km \: s}^{-1}$ flow (lower panel). We also show the time evolution of the fraction of the total mass of carbon 
that is  in the form of C$^{+}$ (green dashed line), C (orange dash-dotted line) and CO (blue dot-dot-dot-dashed line).
In the right-hand column, we show the peak values of the fractional abundances of H$_{2}$ and CO. These are
computed relative to the total number of hydrogen nuclei, and so the maximum fractional abundances of H$_{2}$
and CO are 0.5 and $1.4 \times 10^{-4}$ respectively. Again, we show results for the  6.8 km s$^{-1}$ flow 
in the upper panel and the 13.6 km s$^{-1}$ flow in the lower panel.  Note that the scale of the horizontal axis 
differs between the upper and lower panels.}
\label{fig:chemevol} 
\end{figure*}

An overview of the chemical state of the gas can be found in Figure \ref{fig:chemevol}. The left-hand plots depict how the global chemical state of the gas evolves as the flow advances. They show the fraction of the available hydrogen that is
in the form of H$_{2}$, and the fraction of the available carbon that is in the form of C$^+$, C, or CO. The fraction of the
total carbon that is incorporated into other molecules, such as HCO$^{+}$, is always very small and is not plotted.
The right-hand plots show the maximum abundances of H$_2$ and CO within the simulation, which tells us whether
there are any molecular-dominated regions within the flow. Note that in this plot, the abundances are given with respect 
to the overall number of hydrogen nuclei (a conserved quantity), such that gas in the form of pure H$_2$ will have a
fractional abundance of  H$_2$ that is 0.5, whereas gas in which all of the carbon is in the form of CO will have a fractional
abundance of CO that is $1.4 \times 10^{-4}$.

We start by looking at the evolution of the H$_2$ in the cloud in Figure \ref{fig:chemevol}. The left-hand plot shows that the gas goes from being completely atomic -- as in our initial conditions -- to having around 10 percent of its hydrogen in molecular
form by the point at which star formation sets in ($\sim 7$ percent in the case of the slow flow, and $\sim 12$ percent in the case of the fast flow).  The initial rise in the amount of H$_2$ is also sharp, going from essentially zero to around a percent over a period of less than 2~Myr in each flow. Such a rapid rise can be understood by looking at the density evolution in Figure \ref{fig:rhotemp}. We see that for each flow, the sudden rise in the H$_2$ fraction is accompanied by a rapid rise in the amount of gas with a density above 100 cm$^{-3}$. Since the formation time of H$_2$ is of the order of $10^9 / n$~Myr \citep{hm79},
where $n$ is the number density of the gas, we see that once the gas density exceeds 100 cm$^{-3}$, the time required
to convert a large fraction of the hydrogen to molecular form becomes of the order of a few Myr. Therefore, the sudden
appearance of H$_2$ is simply a consequence of the structure that is formed in the flows. 

Figure \ref{fig:chemevol} also shows that some pockets of gas can become almost fully molecular very early in the calculation, as shown the right-hand plots of Figure \ref{fig:chemevol}. Again this is simply a reflection of the density evolution that we see in Figure \ref{fig:rhotemp}. What is interesting is that these pockets of H$_2$  appear very early in the flows' evolution, well before the onset of star formation. In the slow flow, the pockets of molecular gas appear more than 10 Myr before the first star-forming core, and even in the fast flow, these pockets precede the star formation by about 3 Myr. In both cases, the regions of molecular hydrogen appear long before the flows have been able to assemble anything that could be construed as a star-forming cloud. As such, they exist in relative isolation during the pre-assembly phase. 

In contrast to the early appearance of H$_2$, the appearance of CO occurs extremely late. The simulations start with all of their carbon in the form of C$^+$, and we see that most of the carbon stays in this form as the flows evolve, even once star formation has begun within the dense gas.  Very shortly after the start of the simulation, a small fraction of the C$^{+}$ recombines, yielding neutral atomic carbon, and the amount of this that is present in the flow rises over time as the amount of cold, dense gas increases. In particular, the same increase in density that is seen in Figure~\ref{fig:rhotminmax} also causes the amount of neutral carbon in the flows to rise by an order of magnitude within only 1--2~Myrs.

The same change in the cloud's structure that causes this increase in the abundance of neutral carbon and that is responsible for the sharp rise in the H$_{2}$ fraction also prompts a rise in the overall CO fraction (left-hand panels of Figure~\ref{fig:chemevol}). However, the CO abundance still remains extremely small in comparison to the C or C$^{+}$ abundances:
less than one-millionth of the total amount of carbon has been converted into CO at this point in the evolution of the flows.
From this point on, the CO fraction steadily increases until the onset of star formation. However, even then, CO is about an order of magnitude less abundant than neutral atomic carbon.

The evolution of the peak CO abundance in the cloud also differs markedly from the peak H$_2$ fraction. We see (from the right-hand panels in Figure \ref{fig:chemevol}), that the first region with a  CO fraction high enough to be regarded as `molecular', only appears in the final Myr before the onset of star formation. If we look back to Figures \ref{fig:rhotminmax} and \ref{fig:rhotemp} we see that in each of the flows, the sharp rise in the CO abundance occurs at the same time as the appearance of gas with a number density $n > 10^{4} \: {\rm cm^{-3}}$ -- the density typically associated with pre-stellar cores.
Looking more closely at the spatial distribution of CO, we find that this is not a coincidence: the regions in which the carbon is almost completely converted into CO are indeed the same self-gravitating cores that then collapse to form stars (as will be discussed later in Section \ref{sec:coevol}).

We thus see that the question of when we first form a `molecular' cloud depends on which molecule we are talking about. 
Regions with high molecular hydrogen fractions are produced early in the evolution of the flows and can potentially persist for a long time prior to the onset of star formation if the build-up of the cloud occurs slowly. On the other hand, high CO fractions are produced only within self-gravitating gas, and hence occur only around 1~Myr before the onset of star formation, regardless of the details of the flow. Most importantly, the gas is always `fully molecular' by the time it forms stars.

\subsection{Comparison to post-processed CO abundances}

Prior to this work, the only numerical study of the formation of CO in three-dimensional colliding flow models was that carried out by HH08, in which they used a post-processing treatment to determine which regions of their simulated flow would have high CO fractions. They assumed that in order for gas to have a high CO fraction, it would have to satisfy two criteria: it must have a temperature below 50~K and a mean visual extinction $A_{\rm V, mean} \geq 1$. They found that in their models,
CO formation was triggered primarily by the global gravitational collapse of the cloud in a direction perpendicular to the
inflow, and that this resulted in relatively large regions of the cloud developing high CO abundances. They also found that
these high CO abundances preceded the onset of star formation by around 4--5~Myr. In this section, we examine how the
CO fractions predicted by their post-processing algorithm compare with those we obtain from our time-dependent chemical
network, and thereby test whether the assumptions that HH08 use in their study are valid.

\begin{figure*}
\centerline{
  \includegraphics[width=3.0in]{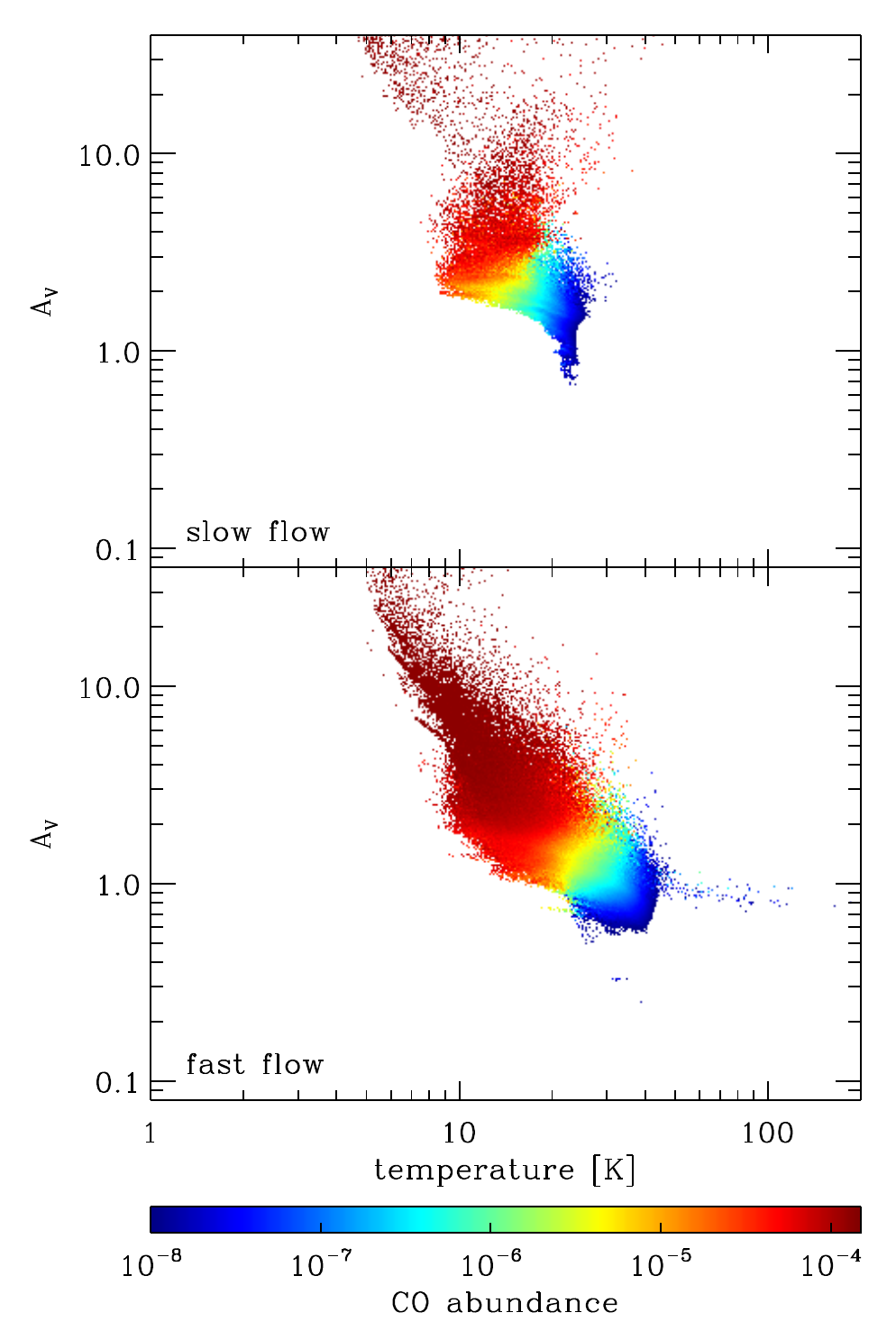}
  \includegraphics[width=3.0in]{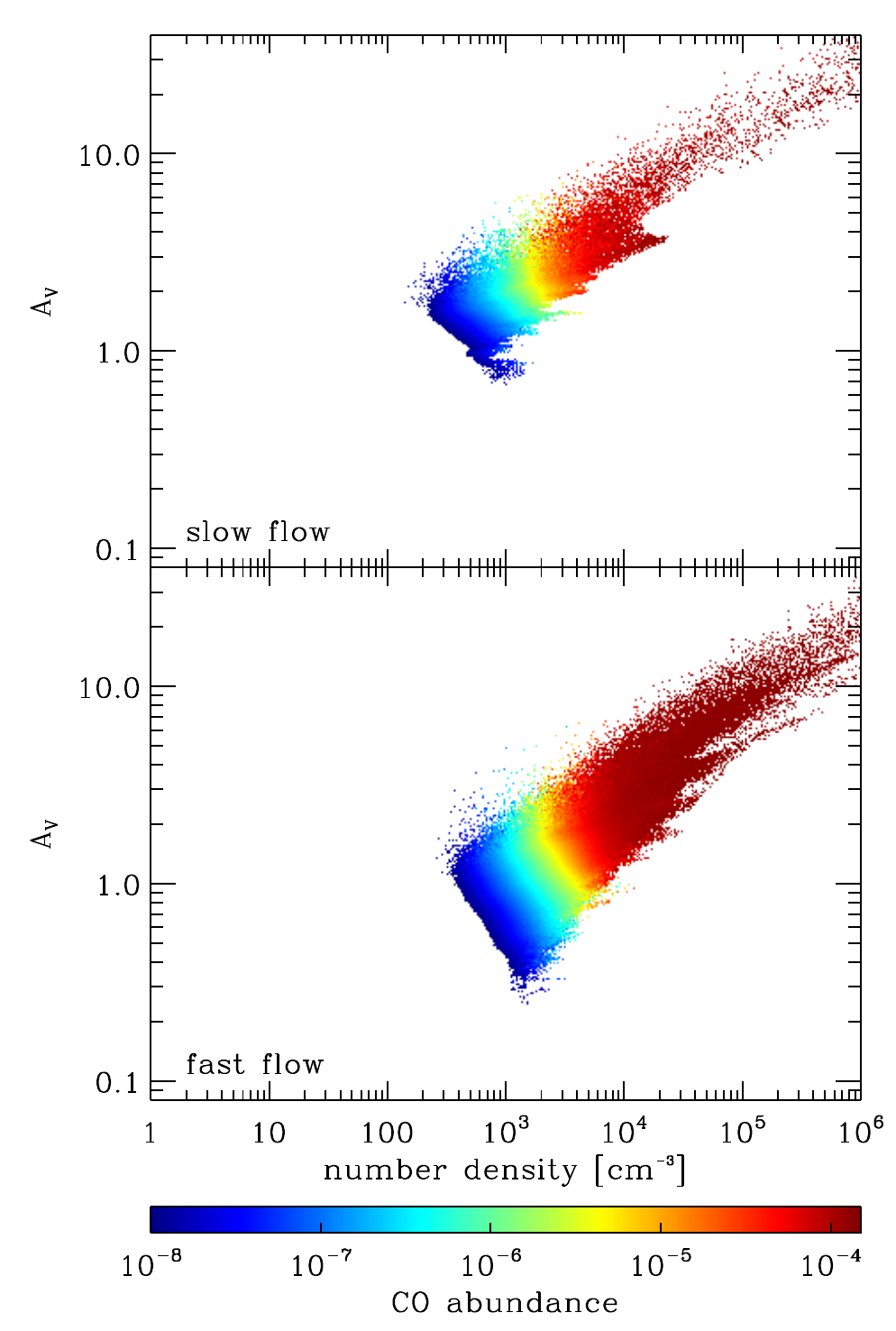}
}
\caption{Weighted mean of the visual extinction seen by each SPH
particle, plotted as a function of temperature (left), and density (right). The colour denotes the fractional abundance of CO, relative to the number of hydrogen nuclei. An SPH particle with all its carbon in the form of CO, would thus have an abundance of $1.4 \times 10^{-4}$. Note that most of the cloud has a CO abundance significantly below $1 \times 10^{-8}$, and so is not visible on this plot, due to the scaling. In these plots we have taken the particle data at a point 0.8 Myr after the onset of the star formation, to give the cloud time to assemble. The visual extinction for each particle is computed via the column densities obtained by the \textsc{treecol} algorithm (Clark et al. 2012), as described in the text.}
\label{fig:temp_av_co} 
\end{figure*}

In our calculations, the visual extinction is calculated during the tree-walk to get the gravitational forces, using our recently developed {\sc TreeCol} algorithm \citep{cgk12}, as described in Section~\ref{sec:numerics}. 
This yields a 48 pixel map of the column densities (and hence visual extinctions) seen by each SPH particle. We use this
map to compute a mean visual extinction  $A_{\rm V, mean}$ for each particle, using the following expression:
\begin{equation}
A_{\rm V, mean} = -\frac{1}{2.5} \ln \left[ \frac{1}{48} \sum_{i=1}^{48} \exp \left(-2.5 A_{\rm V, i} \right) \right],
\end{equation}
where $A_{\rm V, i}$ is the visual extinction associated with pixel number $i$, and we sum over all 48 pixels. 
The weighted mean that we calculate in this fashion accounts for the fact that the photodissociation rates of molecules
such as CO and the photoelectric heating rate of the gas all depend on exponential functions of $A_{\rm V}$, rather
than directly on $A_{\rm V}$ itself. 

In Figure~\ref{fig:temp_av_co} we show the distributions of temperature and mean visual extinction for the particles in both flows, at point 0.8 Myr after the onset of star formation (recall that star formation occurs at a time of 16 Myr in the slow flow, and 4.4 Myr in the fast flow). The points are coloured by the CO abundance of the corresponding SPH particle, and results are only shown for SPH particles with fractional CO abundances greater than $10^{-8}$. This means that very little of the warm gas in the flows appears in the plot, since this material typically has a CO abundance below this threshold. In the case of the slow flow, we see from the figure that high CO fractions are present only in gas with a mean visual extinction $A_{\rm V, mean} > 2$ and a temperature lower than 20~K. In this case, adopting the higher temperature threshold and lower extinction threshold used by HH08 would lead to an overestimate of the the  fraction of the flow that would be traced by CO, and thus identified as a molecular cloud. In the fast flow, the HH08 approach fares better. In this case, the transition from low to high CO abundances does indeed occur for $A_{\rm V, mean} \sim 1$, although for gas with a temperature below around 20--30~K; gas with $T \sim 50$~K and $A_{\rm V, mean} \sim 1$ has a very low CO abundance.

The motivation for the extinction threshold adopted in HH08 was provided by the study of \citet{db88}, who demonstrated 
that in uniform density, plane-parallel cloud models, CO is abundant in regions that have visual extinctions $A_{\rm V} > 1$.
However, the \citet{db88} study did not account for density inhomogeneities in the gas, which have the effect of significantly
complicating the relationship between the CO abundance and the visual extinction \citep[see e.g.][]{g10}. The HH08 approach therefore implicitly assumes that all of the gas with $A_{\rm V, mean} \ga 1$ and $T < 50$~K is dense enough to
sustain a high CO abundance. As we see in Figure~\ref{fig:temp_av_co}, this is not the case in our model clouds: the scatter in the relationship between $A_{\rm V, mean}$ and $n$ is quite large, and so much of the gas with $A_{\rm V, mean} \sim 1$ has a density below $1000 \: {\rm cm^{-3}}$ and hence although it is relatively cold ($T < 50 \: {\rm K}$), it is nevertheless too diffuse to have a high CO abundance. A similar density threshold of around $1000 \: {\rm cm^{-3}}$ is also found in more idealised turbulent-cloud models \citep[see e.g.][]{molina11}.

\begin{figure}
  \includegraphics[width=3.3in]{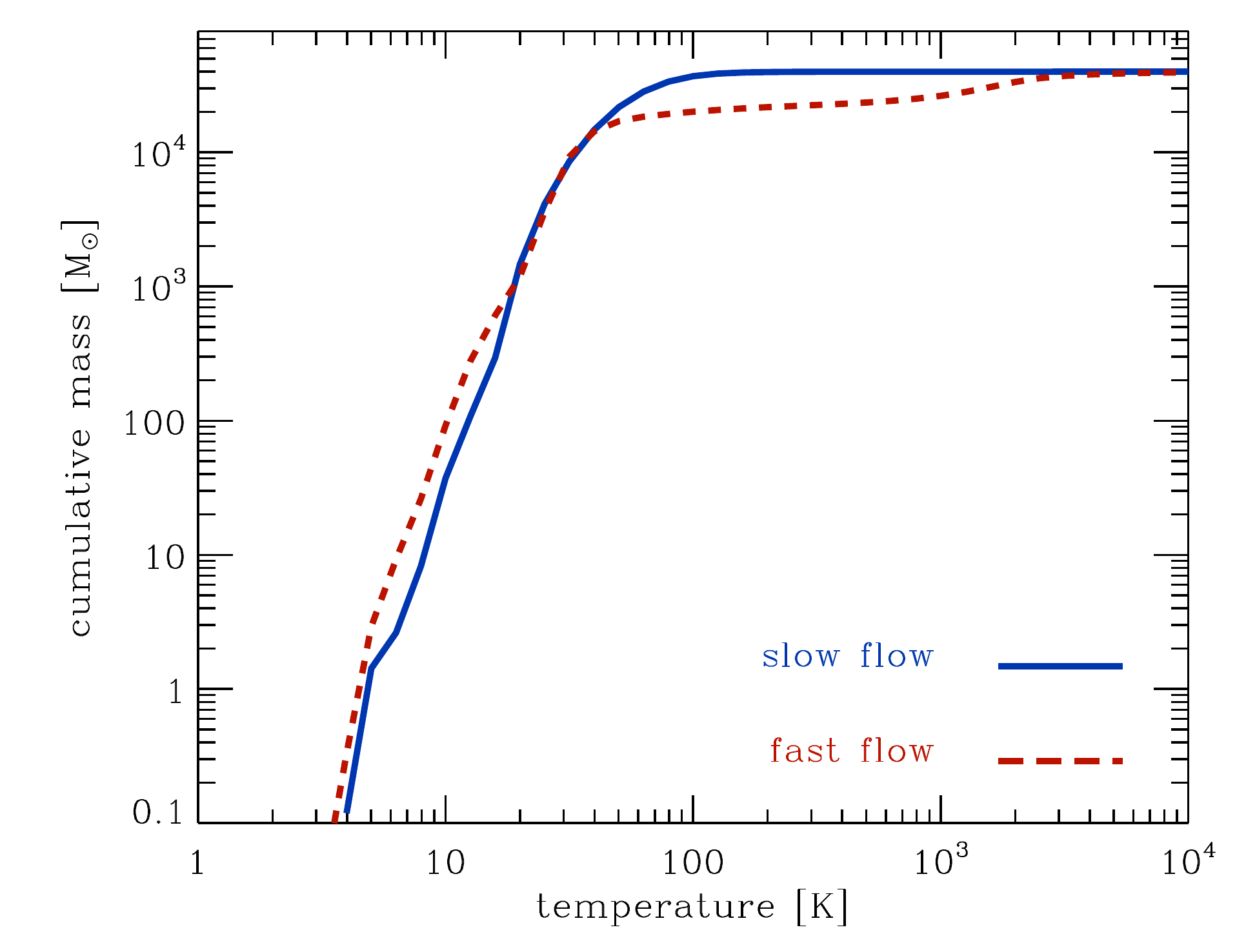}
\caption{The cumulative temperature distribution in the slow and fast flows at the onset of star formation.}
\label{fig:cumutemp} 
\end{figure}

One can see by just how much the true mass of the clouds would be overestimated by looking at Figure \ref{fig:cumutemp}, which shows the cumulative temperature distribution in terms of the flow mass. We see that there is around an order of magnitude more gas with a temperature in the range $20 < T < 50$~K than has a temperature $T < 20$~K. Given that the 
mean visual extinction of gas with $T < 50 \: {\rm K}$ is generally quite high, as shown in Figure \ref{fig:temp_av_co}, we
see that the HH08 approach will typically result in inferred `molecular clouds' that are about 10 times more massive than those predicted by our more self-consistent approach. 

Nevertheless, it is clear that the basic motivation underlying the HH08 approach is sound. Our results suggest that instead
of using a threshold of $T < 50$~K and $A_{\rm V, mean} > 1$, the values $T < 20$\,K and $A_{\rm V, mean} > 2$ would 
give a more accurate picture of the location of the CO-bright gas, at least in the early stages of cloud assembly that we focus on here.

\subsection{Detection via CO emission}
\label{sec:coevol}

The plots in Figure \ref{fig:chemevol} show us that CO becomes highly abundant in some regions of the cloud shortly before
the onset of star formation. However, this on its own does not tell us whether these regions would actually be detectable in
CO emission. In this section, we post-process our results using the RADMC-3D\footnote{http://www.ita.uni-heidelberg.de/$\sim$dullemond/software/radmc-3d/} radiative transfer code in order to construct maps of the $^{12}$CO (1-0) line emission coming from the region immediately surrounding the first star-forming core in each flow. To compute the CO level populations, we use the large velocity gradient approximation \citep{Sobolev57}, as implemented in RADMC-3D by \citet{Shetty2011a, Shetty2011b}. The SPH data is interpolated onto a 256$^3$ grid of size 6.48 pc, centred on the star-forming core. In this manner, the self-consistently computed abundances of the CO from the simulations can be used as the basis of the radiative transfer calculation. The radiative transfer is done along the $x$-axis of the grid, such that the rays are fired from negative $x$ to positive $x$. The resulting maps are hence those that would be seen by an observer sitting at positive $x$ (i.e.\ one who is looking along the flow).

\begin{figure*}
\centerline{
  \includegraphics[width=3.4in]{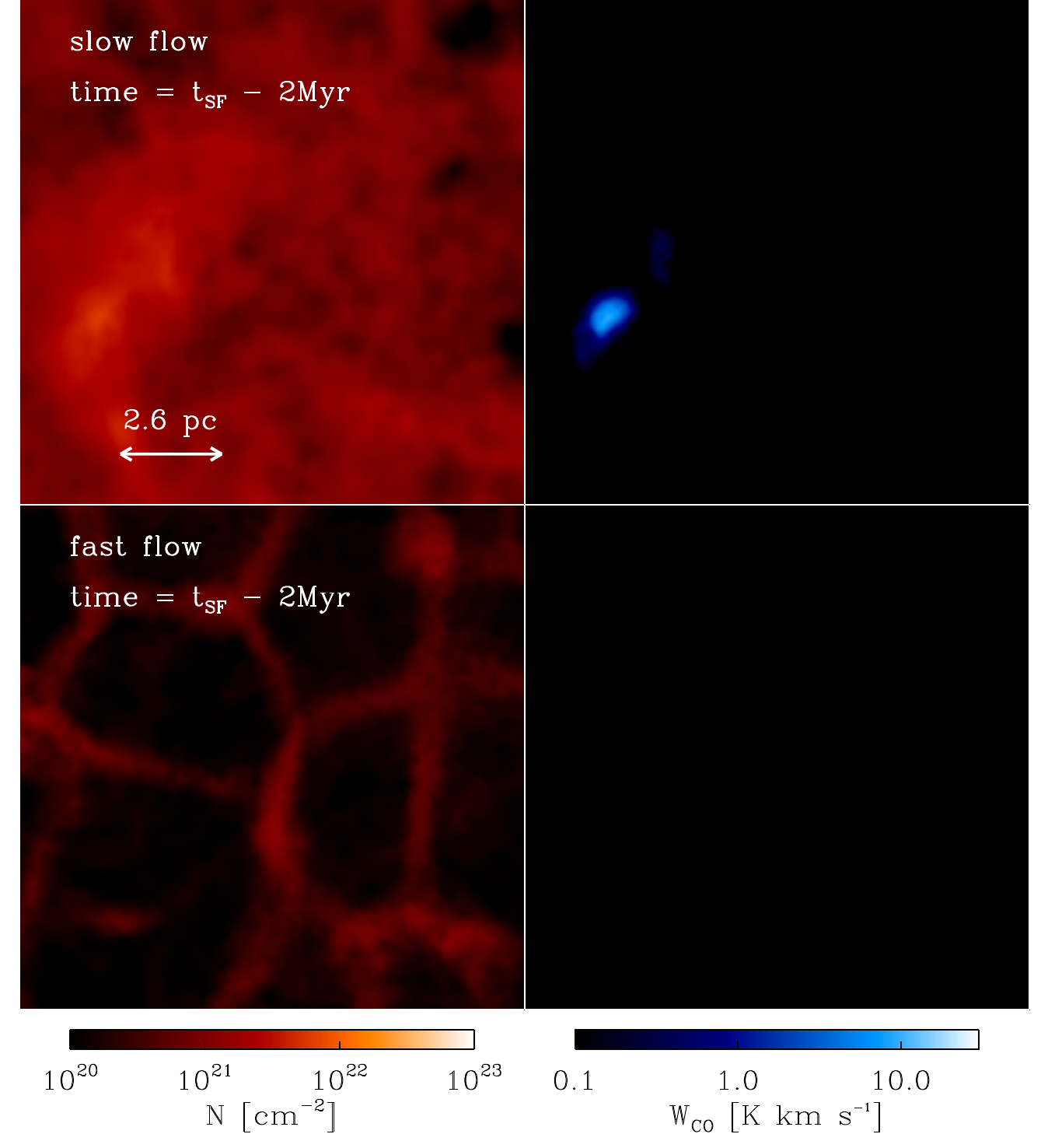}  
  \includegraphics[width=3.4in]{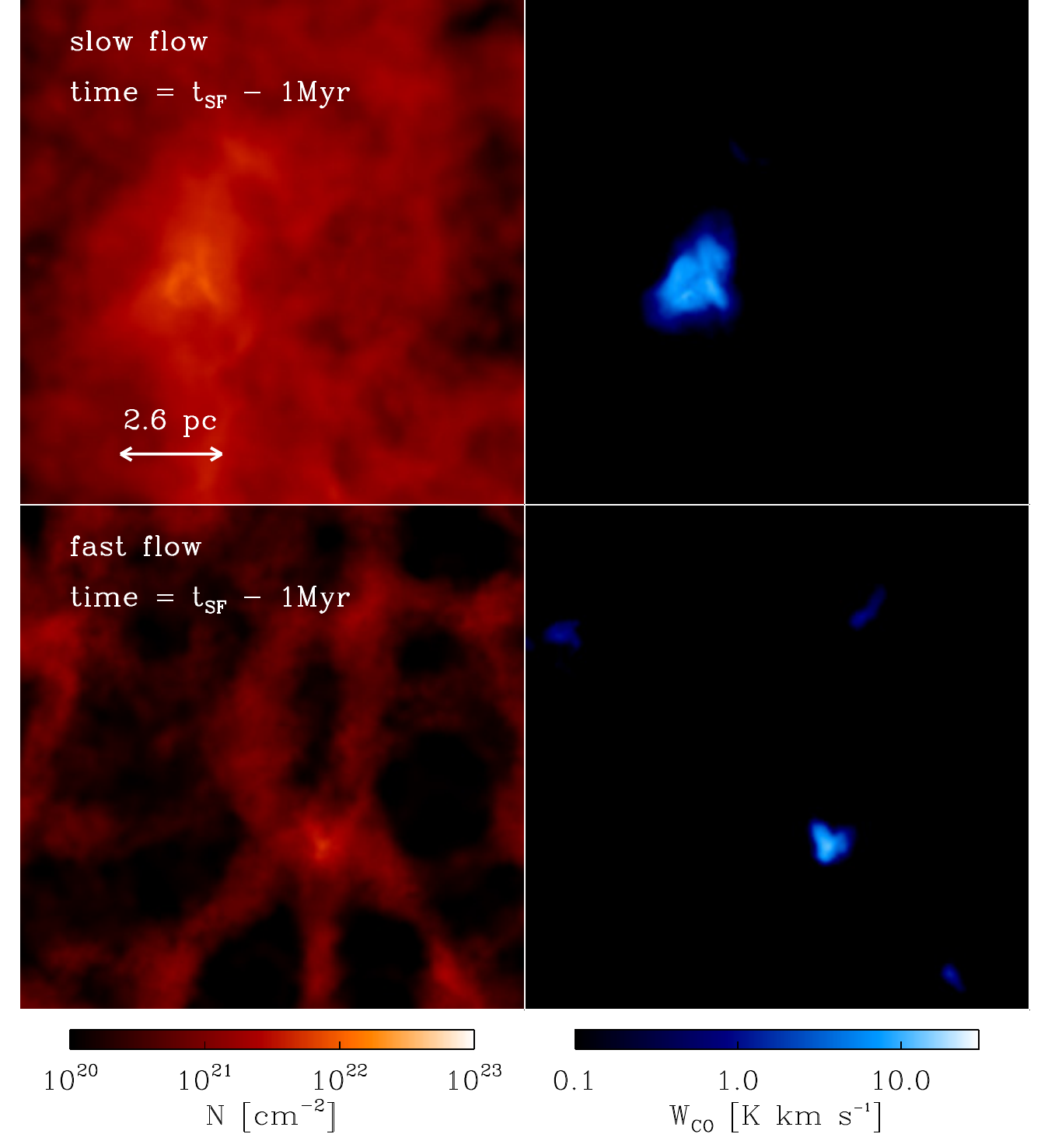}
}
\centerline{ 
  \includegraphics[width=3.4in]{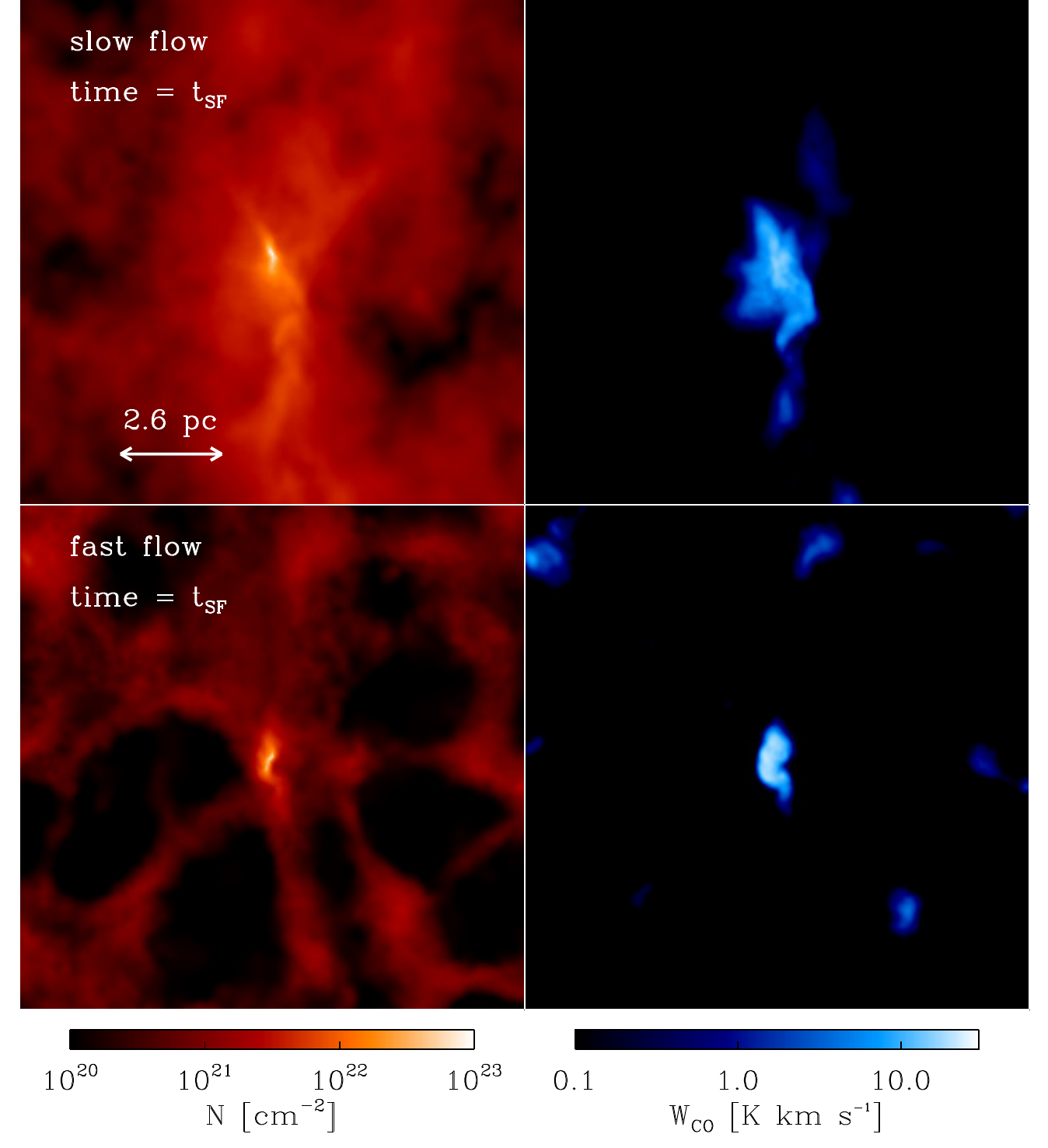} 
  \includegraphics[width=3.4in]{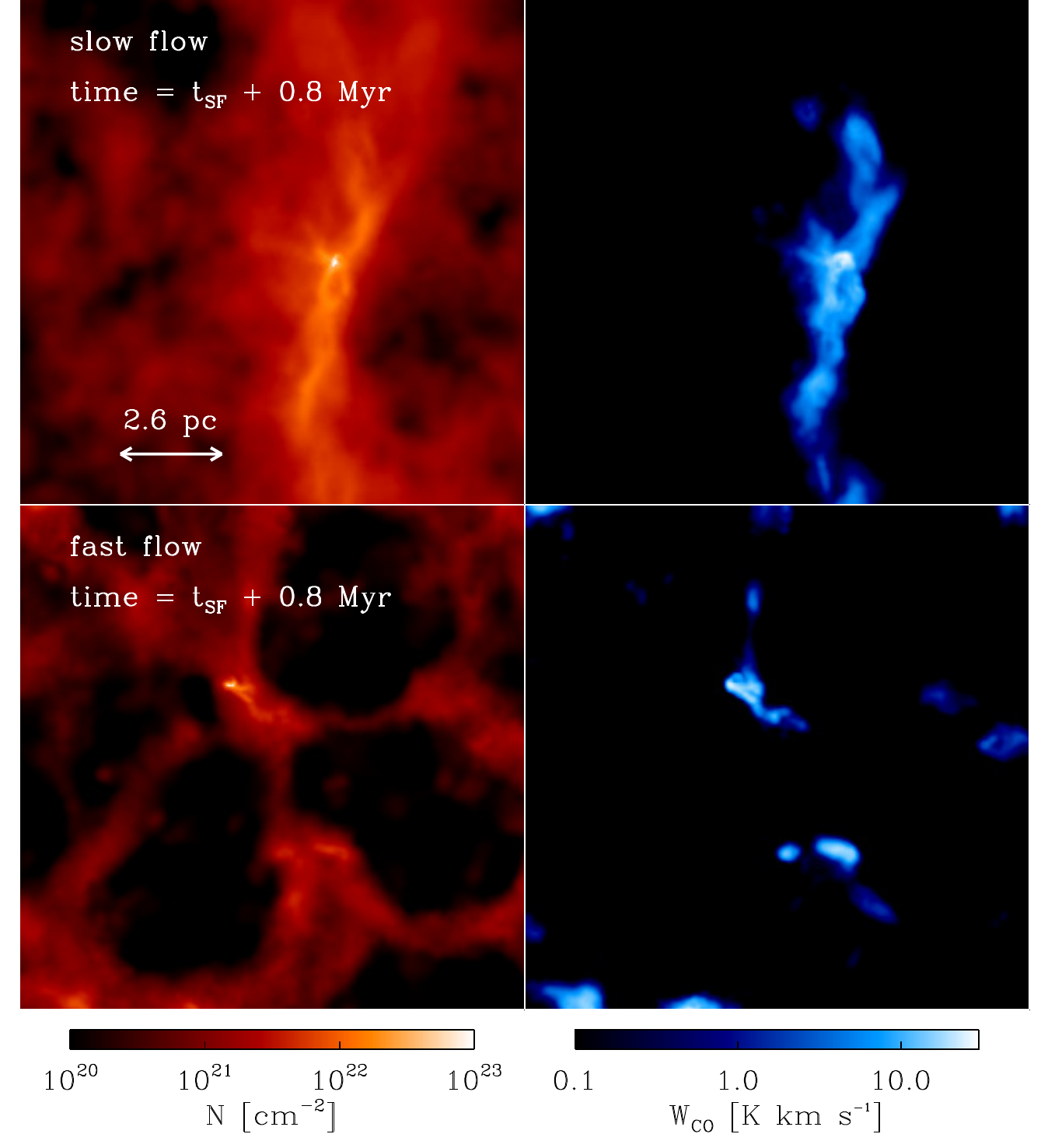}
}
\caption{The images show the evolution of the column number density, $N$, and the velocity-integrated intensity in the
$J = 1 \rightarrow 0$ line of $^{12}$CO,  $W_{\rm CO}$ (1-0), for the region in which the first star forms in each of the flows. Four times are shown: 2 Myr prior to star formation (upper left-hand panels), 1 Myr prior to star formation (upper right-hand panels), the point of star formation (lower left-hand panels), and 0.8 Myr after the onset of star formation (lower right-hand panels). The CO integrated intensity map is obtained via a radiative transfer calculation performed with the 
{\textsc RADMC-3D} code, and uses the large velocity gradient approximation to compute the CO level populations.}
\label{fig:wcoevol} 
\end{figure*}

The images in Figure \ref{fig:wcoevol} show the results from the radiative transfer calculations, along with the column number density for the same region. The line transfer data is shown as the velocity-integrated intensity ($W_{\rm CO}$) -- the quantity that is commonly used to determine H$_{2}$ column densities via the so-called `X-factor' \citep{Solomon1987,Young1991, Dame2001}. We perform the radiative transfer at 4 times in each flow's history, spanning a period from 2 Myr before the onset of star formation, to 0.8 Myr after (which is also the point at which we halt the simulations due to the rapidly increasing computational expense).

We see that in both cases, $W_{\rm CO}$ is large in the region immediately surrounding the star-forming core, and that the dense prestellar core is embedded in a larger region of diffuse CO emission. The range of values for $W_{\rm CO}$ in this
region -- from a few K km\,s$^{-1}$ to around 20-30 K km\,s$^{-1}$ -- appears to be consistent with the range of values found in local clouds such as the Taurus molecular cloud \citep{goldsmith2008, narayanan2008}. However, the length-scale associated with the CO-bright regions does differ between the two flows. For the slow flow, the region around the prestellar core is 2~pc in length at the onset of star formation, while in the fast flow the region is somewhat smaller, having a diameter 
of roughly 0.5~pc. In both cases, the extent of the CO emission reflects the column density distribution in the gas. In general,  however, it seems that at these early times the CO is confined to regions that are undergoing gravitational collapse. 

The images shown in Figure \ref{fig:wcoevol} also demonstrate that both the size of the CO-bright region and the strength of the emission from this region vary strongly with time. At a time of 2~Myr before the onset of star formation, there is essentially no visible CO emission coming from the fast flow, and only one small region of relatively faint emission in the slow flow.\footnote{Strictly speaking, there is a very low level of CO emission coming from every part of the cloud, but in the image we
only indicate the emission when $W_{\rm CO} \geq 0.1 \: {\rm K} \: {\rm km} \: {\rm s^{-1}}$, as emission below this threshold would not in practice be detectable.} However, the size of the CO-bright region and the strength of the emission from this
region both increase rapidly as the prestellar core begins to undergo gravitational collapse. From the column density images, we can see that this change largely just reflects the change in the column density distribution that occurs as the core collapses. Although the observational definition of `detection' of a molecular cloud via CO emission can vary, depending on the method used to extract the cloud from the data-set and the telescope used to conduct the survey, a value of $W_{\rm CO} \ga 1 \: {\rm K} \: {\rm km} \: {\rm s^{-1}}$ appears to be a practical threshold for the robust detection of `molecular gas' (see, for example, the discussions in \citealt{goldsmith2008} and \citealt{heyer2009}). In our maps, we see that  $W_{\rm CO}$ falls off rapidly below this value, and so we can conclude that the emission seen in Figure \ref{fig:wcoevol} is detectable. The $W_{\rm CO}$ maps therefore confirm the picture that we discussed in Section \ref{sec:chem}, namely that the emergence of CO occurs fairly late in the evolution of the flow, and at roughly 2 Myr before the onset of star formation.

From this we can conclude that the colliding flow model can indeed produce an observationally detectable molecular cloud 
prior to the point at which star formation begins in the cloud. Furthermore, the timescale over which the CO becomes observable is typically around 2 Myr, and appears to have little dependence on the speed of the flow that is forming the 
cloud. 

\section{The onset of star formation}
\label{tsf}

Our discussion so far has touched upon conditions in the clouds that accompany the onset of star formation. In this section we take a closer look at how the gravitational instability occurs and compare our results with previous studies of self-gravitating colliding flows. 

\begin{figure*}
\centerline{
  \includegraphics[width=2.9in]{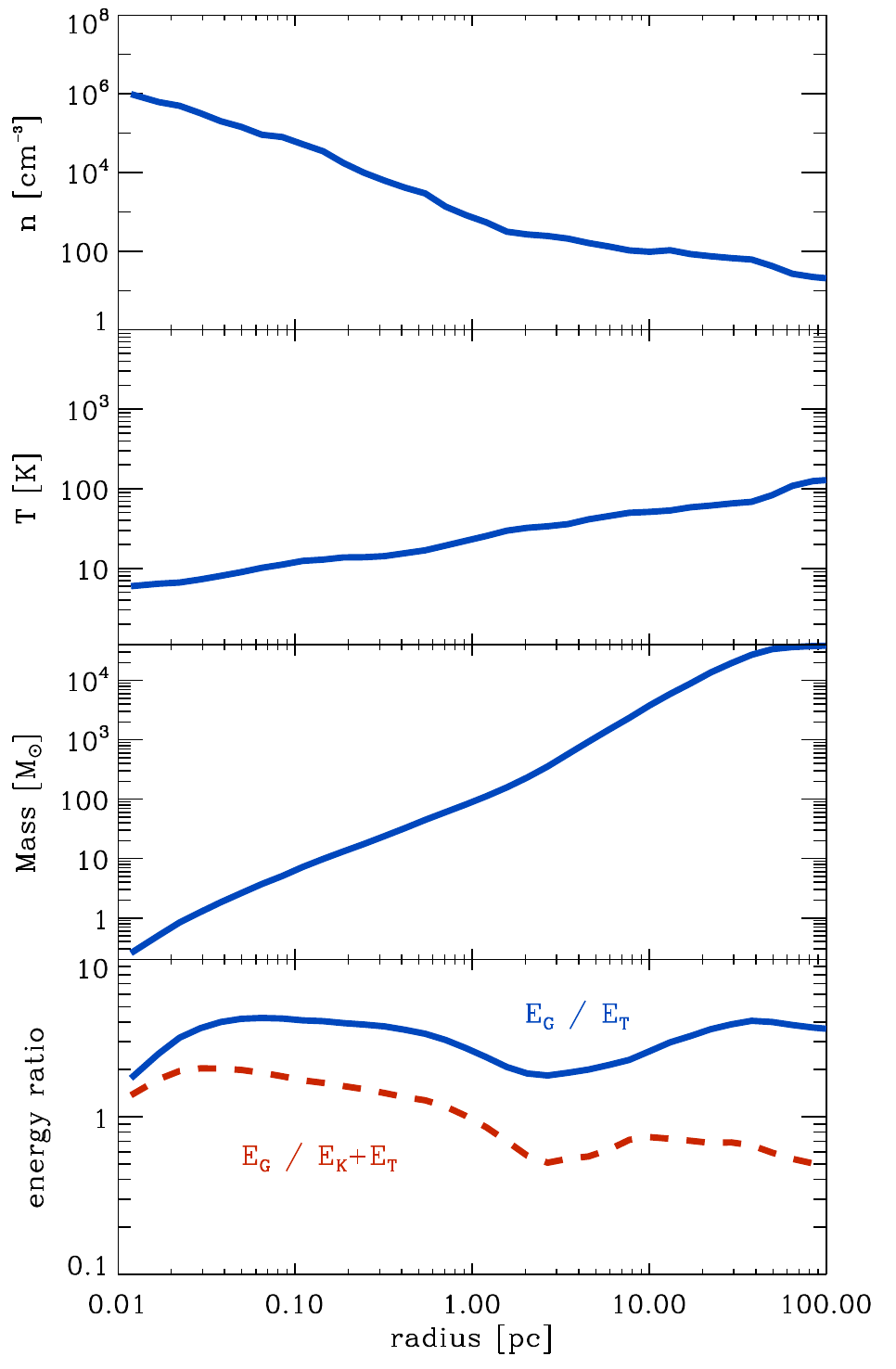}
  \includegraphics[width=2.9in]{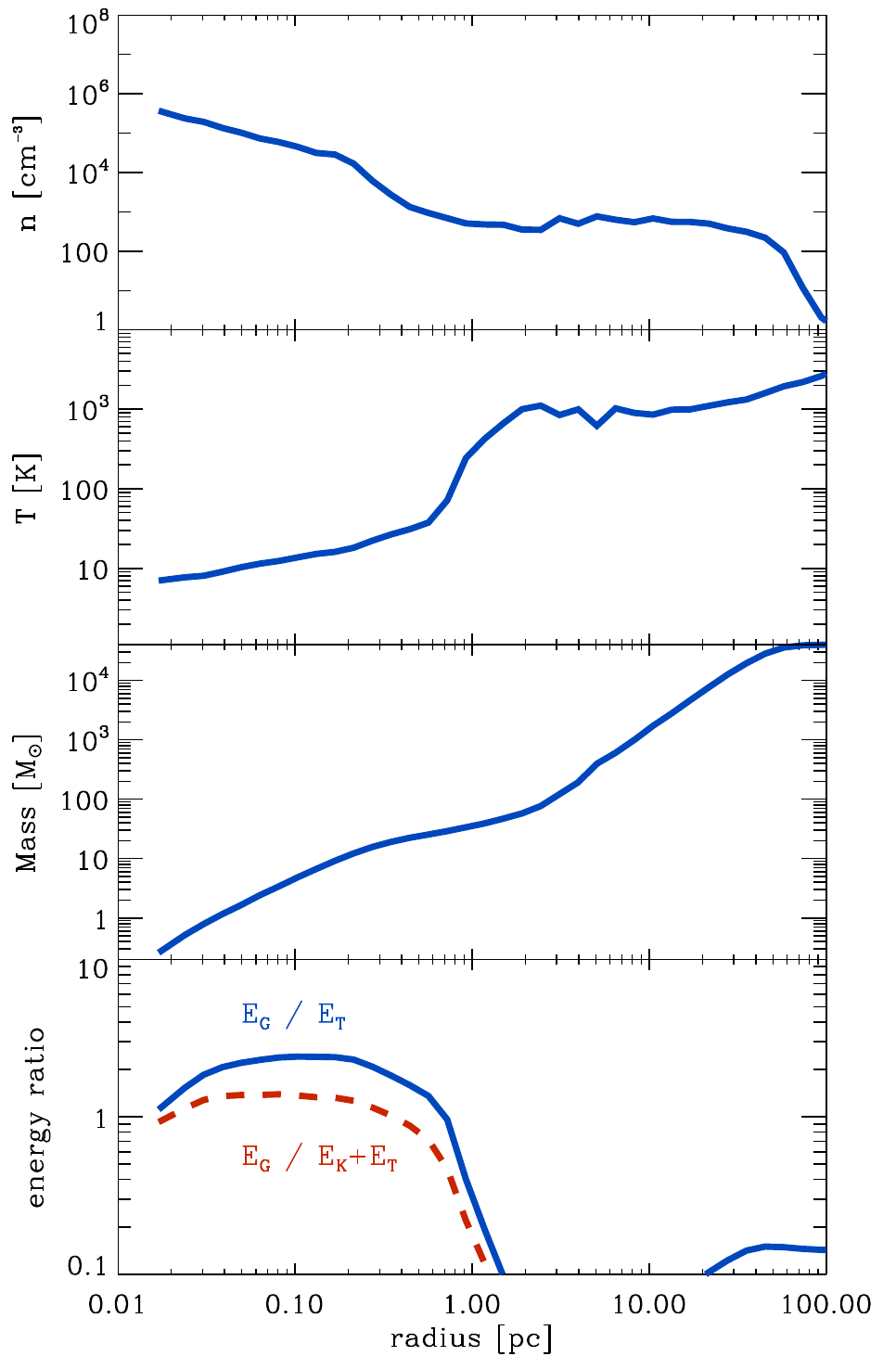}
}
\caption{The panels show the radial profiles centered on the first SPH particle to find itself in a star in the 6.8 km s$^{-1}$ flow simulation (left) and in the 13.6 km s$^{-1}$ flow (right). The profiles are computed just before the onset of star formation in each case. Temperature and density are mass-weighted averages, while `mass' denotes the mass enclosed at a given radius. In the bottom panel, $E_{\rm G}$, $E_{\rm K}$ and $E_{\rm T}$ denote the magnitude of the gravitational energy, the kinetic energy, and the thermal energy, respectively.}
\label{fig:radeng} 
\end{figure*}

The plots in Figure \ref{fig:radeng} show how the properties of the gas vary as a function of distance from the first collapsing
core to form in each simulation. We plot the radially averaged number density and temperature of the gas, along with the mass enclosed within a sphere of that radius. We also examine how the ratios $E_{\rm G} / E_{\rm T}$ and $E_{\rm G} / (E_{\rm K} + E_{\rm T})$ vary with distance from the core, where $E_{\rm G}$ is the magnitude of the gravitational energy, 
$E_{\rm T}$ is the thermal energy and $E_{\rm K}$ is the kinetic energy. Results for the slow flow are shown on the left,
and those for the fast flow are shown on the right. The profiles are taken from the last output snapshot produced by each
simulation before the onset of star formation, and hence correspond to a time less than 10$^5$ years prior to the formation 
of the first sink particle.\footnote{We produce output snapshots every $10^{5} \: {\rm yr}$ during the runs.}

As in the previous sections, we again see a very different picture when we compare the conditions surrounding the star-forming core in each flow. The main difference between the simulations is that the slow flow is almost entirely Jeans unstable by the onset of star formation (that is, its gravitational energy is greater than its thermal energy), while the fast flow is in general gravitationally stable. This is due primarily to the fact that turbulence in the fast flow has not had time to trigger the cooling instability, and so most of the gas is still in the warm phase. This is apparent when we compare the radial temperature profiles in the each flow; the temperature rises gradually in the slow flow, while in the fast flow we see a sudden rise in the temperature at around 1 pc from the centre of the core. This sudden temperature rise marks the boundary of the collapsing region in the fast flow. In the slow flow, we see instead that this boundary is marked by a rise in the kinetic energy, indicating that it is the turbulent motions in the flow that determine which regions can locally collapse and which cannot.

The fact that the entire cloud becomes gravitationally unstable in the slow flow is broadly consistent with the previous numerical studies of self-gravitating colliding flows \citep{enrique2006,Heitsch2006,vs07,Hennebelle2008,banerjee09,enrique2009,rg10}. Typically it is found that the first star-forming cores appear while the post-shock layer itself is undergoing collapse (see the discussions in \citealt{hhb2008} and \citealt{enrique2009}). In contrast, the behaviour seen in our fast flow seems to differ from the standard picture, in that the collapsing regions are localised, and not accompanied by the collapse of the post-shock layer. The origin of this difference is unclear, and the fact that the existing studies (at least the high resolution, self-gravitating studies) do not present data similar to that expressed in our Figure \ref{fig:rhotemp}, means that a detailed investigation of this difference lies beyond the scope of this paper. However one potential difference between our results and the previous studies is that we include a more self-consistent treatment of the photoelectric emission heating, as we are able to accurately calculate the attenuation of the external radiation field via our {\sc TreeCol} algorithm. Thus it is likely that our fraction of very cold gas (i.e.\ gas colder
than 30~K) is much higher, allowing gravitationally bound structures  to form more readily in our simulations.

A final interesting feature regarding the onset of star formation in our simulations is the fact that both flows have roughly the same mass in the local collapsing region -- at least to within a factor of two. Given the very different conditions in the flow at the point where the gravitational instability sets in, this is perhaps somewhat surprising. However, the temperature of the gas in the cold regime is a strong function of density, as can been seen in Figure \ref{fig:rhotphase}. Although the fast flow does have slightly higher temperatures on average at a number density of $10^3$ cm$^{-3}$ -- the density at which the gravitational instability sets in, as shown in Figure \ref{fig:radeng} -- the difference is fairly small. As such, the Jeans mass in the collapsing regions is roughly the same. This ability of the gas to regulate its temperature is obviously crucial for explaining why the conditions in regions of nearby star formation are so similar \citep{larson85, larson05, elmegreen2008}.

\section{Discussion}
\label{discuss}

The results of this study show that it is possible to assemble star-forming regions in two ways -- either by putting together pre-existing, cold molecular clumps, or forming them directly from the warm ISM via the cooling instability. The former scenario is what we see in our `slow' calculation, in which the majority of the gas in the incoming flow has time to cool before hitting the shock at the edge of the central cloud. As such, the gas is cold and clumpy by the time it encounters material coming from the other direction, and thus the set-up is more akin to the well-studied `cloud-cloud' collision scenario rather than the classic colliding flow set-up, in which only the interface between the flows has been perturbed. In fact, our set-up is similar to that employed by \citet{rg10}, in which they show that the exact properties of the star formation in the clouds are sensitive to the initial level of turbulence that is injected into the colliding flows. 

The second scenario, in which the dense bound regions form directly from the cooling instability in the shocked region, is  represented by our `fast' flow calculation. In this case, although the flow is perturbed throughout, there is insufficient time for the thermal instability to occur in the low density inflowing gas and so the majority of the gas entering the shock is still in the 
unstable or warm phases ($T > 1000$~K). In this case, the warm, incoming gas cools and forms a region that grows until it is gravitationally unstable. This behaviour is similar to that seen in the simulations of \citet{banerjee09}, although as discussed above in Section \ref{tsf}, we do not see the accompanying collapse of the post-shock layer in this case. As the temperature in the post-shock gas is set by the balance between photo-electric emission and C$^+$ line-cooling, it is similar to that found in the star-forming region that forms in the slow flow. Therefore, although the assembly process is very different in the two cases, the resulting star-forming regions have similar properties.

Which of these two scenarios is more likely to occur in reality? \citet{Heitsch2006} point out that the density regime from 1 to 10 cm$^{-3}$ is thermally unstable, and so it is likely that clouds are at least in part assembled via collisions between cold gas. The conditions that we see in our fast flow are therefore likely to be somewhat artificial, as in reality the flows would have had ample time to undergo the thermal instability. Our calculations also suggest that much of the cold gas that participates in the cloud formation process may already consist of H$_{2}$ long before the final `cloud' (or star-forming region) is assembled. 
This gas would not be observable via its CO emission until it has become part of the main cloud. As such, our simulations support the picture outlined by \citet{pal01}, who propose that clouds form out of pre-existing, CO-dark, molecular gas. Similar behaviour was also suggested by the study by \citet{dobbs08}, which looked at GMC formation in spiral arms.
In this context, it is interesting to note that there is growing observational evidence for the existence of significant quantities of H$_{2}$-rich, CO-poor gas in the Milky Way \citep{gren05,langer10,pineda10,velu10,ade11}. Our simulations suggest that this so-called ``dark'' molecular gas is a natural outcome of the process of cloud assembly.

Finally, we note that as the CO appears fairly late in the cloud assembly process, it plays only a minor role in the thermodynamics of cloud formation. Throughout much of the assembly of the cloud, the cooling is dominated  by
atomic line cooling from C$^+$ and O, with CO becoming important only in regions where the gas is already cold
and dense. This is consistent with the picture presented in \citet{gc12a}, where it was argued that molecules are not 
needed to form stars, but rather accompany the star formation process.

\section{Conclusions}
\label{conc}

We have presented results from the first three-dimensional calculations to follow the non-equilibrium chemistry of H$_2$ and CO formation from the warm neutral medium to the cold molecular regime, in the context of the colliding flow picture of molecular cloud formation \citep{elmegreen93}. Our time-dependent chemical network is coupled to a detailed model of the thermodynamics of the ISM that accounts for the main heating and cooling processes, such as atomic and molecular line cooling (from H, C$^+$, C, O, CO and H$_2$), photoelectric emission and cosmic ray heating, and the energy transfer between the gas and dust. The chemical and thermodynamical model are incorporated into the publicly available smoothed particle hydrodynamics code {\sc Gadget2} \citep{springel05}.

We have presented calculations in this paper that follow the evolution of two colliding flows with velocities 6.8 km s$^{-1}$ and 13.6 km s$^{-1}$, which we refer to as our `slow' and `fast' flows, respectively. The flows start essentially fully atomic in composition, except for a small ionisation fraction of $6.5 \times 10^{-3}$, set by the equilibrium between cosmic ray ionisation and radiative recombination. The density and temperature in both flows is initially 1 cm$^{-3}$ and 5000 K, respectively. To seed the structure that will eventually develop into molecular clouds, we impose a turbulent field with $v_{\rm rms} = 0.2\,c_{\rm s}$, where the isothermal sound speed is 5.6 km s$^{-1}$. The velocities follow the observed scaling law of $v \propto L^{0.5}$ (e.g. \citealt{hb2004}).

Star formation -- as captured by the formation of `sink particles' \citep{bbp95} -- occurs in the calculations at around 16 Myr for the slow flow, and 4.4 Myr for the fast flow. By this point, each flow has over 1000 \solmas of gas sitting below a temperature of 30 K. We find that in both cases, the flows are able to form pockets of gas that have very high H$_2$ fractions, and that this occurs early in the evolution of the flow, at a time of around 6.5 Myr in the slow flow, and 1.5 Myr in the fast flow.

However the formation of regions with a high CO abundance is significantly delayed with respect to the molecular hydrogen. In both flows, we find that significant quantities of CO do not form until around 2~Myr before the onset of star formation.
Using the results from our SPH simulation as input to line radiative transfer calculations, we were able to produce synthetic
emission maps for the CO ($1-0$) line. These synthetic emission maps confirm that the appearance of CO-bright regions 
in the clouds precedes the formation of stars by no more than about 2~Myr.

Our results suggest that even in cases when the assembly of the cloud takes place over many millions of years, the interval between the cloud becoming observationally detectable in CO and the onset of star formation is short, of the order of 1--2~Myr. This is consistent with the idea that molecular clouds can form out of ``dark'' molecular gas \citep{Wolfire2010} -- cold gas that has a high H$_2$ content but contains little or no CO \citep{pal01, dobbs08, dobbs2008}.

\section*{Acknowledgements}
The authors would like to thank Robi Banerjee, Fabian Heitsch, Mordecai-Mark {Mac Low}, Jim Pringle, Rahul Shetty, and Enrique V\'azquez-Semadeni. for many enlightening discussions about the results presented in this paper. PCC, SCOG and RSK acknowledge financial support from the German Bundesministerium f\"ur Bildung und Forschung via the ASTRONET project STAR FORMAT (grant 05A09VHA), the Deutsche Forschungsgemeinschaft (DFG) via SFB 811``The Milky Way System'' (sub-projects B1 and B2), and from the Baden-W\"urttemberg-Stiftung by contract research via the programme {\em Internationale Spitzenforschung II} (grant P-LS-SPII/18) of the Baden-W\"{u}rttemberg Foundation. PCC is supported by grant CL 463/2-1, part of the DFG priority program 1573 ``Physics of the Interstellar Medium''. The simulations described in this paper were performed on the {\em Ranger} cluster at the Texas Advanced Computing Center, using time allocated as part of Teragrid project TG-MCA99S024. Additional calculations were also performed using the {\em kolob} cluster at the University of Heidelberg, which is funded in part by the DFG via Emmy-Noether grant BA 3706, and via a Frontier grant of Heidelberg University, sponsored by the German Excellence Initiative as well as the Baden-W\"urttemberg Foundation.


\begin{thebibliography}{}

\bibitem[Ade et~al.(2011)]{ade11}
Ade, P.~A.~R., et~al. (Planck collaboration), 2011, A\&A, 536, A19

\bibitem[Audit \& Hennebelle(2005)]{ah05}
Audit, E., \& Hennebelle, P.\ 2005, A\&A, 433, 1

\bibitem[Banerjee et~al.(2009)]{banerjee09}
Banerjee, R., V\'{a}zquez-Semadeni, E., Hennebelle, P., \& Klessen, R.~S. 2009, MNRAS, 398, 1082

\bibitem[Ballesteros-Paredes, Hartmann \& V\'azquez-Semadeni(1999)]{bp99}
Ballesteros-Paredes, J., Hartmann, L., \& V\'azquez-Semadeni, E.\ 1999, ApJ, 527, 285

\bibitem[Bate, Bonnell \& Price(1995)]{bbp95}
Bate, M.~R., Bonnell, I.~A., \& Price, N.~M.\ 1995, MNRAS, 277, 362

\bibitem[Bate \& Burkert(1997)]{bb97} Bate, M.~R., \& Burkert, A.\ 1997, MNRAS, 288, 1060 

\bibitem[Benz(1990)]{benz90}
Benz, W., 1990, in `Proceedings of the NATO Advanced Research Workshop on The
Numerical Modelling of Nonlinear Stellar Pulsations Problems and Prospects',
ed.\ J.~R.~Buchler, (Dordrecht: Kluwer), 269

\bibitem[Bergin et~al.(2004)]{berg04}
Bergin, E.~A., Hartmann, L.~W., Raymond, J.~C.,
\& Ballesteros-Paredes, J.\ 2004, ApJ,  612, 921

\bibitem[Bergin \& Tafalla(2007)]{bt07}
Bergin, E.~A., \& Tafalla, M.\ 2007, ARA\&A, 45, 339

%\bibitem[Bigiel et al.(2008)]{bigiel08} 
%Bigiel, F., Leroy, A., Walter, F., Brinks, E., de Blok, W.~J.~G., Madore, B., 
%\& Thornley, M.~D.\ 2008, AJ, 136, 2846

%\bibitem[Bigiel et al.(2011)]{bigiel11}
%Bigiel, F., Leroy, A.~K., Walter, F., Brinks, E., {de Blok}, W.~J.~G., Kramer, C., Rix, H.~W., 
%Schruba, A., Schuster, K.~F., Usero, A., \& Wiesemeyer, H.~W.\ 2011, ApJL, 730, L13

%\bibitem[Black \& Dalgarno(1977)]{bd77}
%Black, J.~H., \& Dalgarno, A. 1977, ApJS,  34, 405

\bibitem[Black(1994)]{bl94}
Black, J.~H. 1994, ASP Conf.\ Ser.\ 58, in The First Symposium on the Infrared Cirrus
and Diffuse Interstellar Clouds, eds.\ R.~M.~Cutri \& W.~B.~Latter, (San Francisco:ASP), 355

\bibitem[Blitz \& Rosolowsky(2004)]{br04}
Blitz, L., \& Rosolowsky, E.\ 2004, ApJ, 612, L29

\bibitem[Blitz \& Rosolowsky(2006)]{br06}
Blitz, L., \& Rosolowsky, E.\ 2006, ApJ, 650, 933

\bibitem[Blitz et~al.(2007)]{blitz07}
Blitz, L., Fukui, Y., Kawamura, A., Leroy, A., Mizuno, N., \& Rosolowsky, E.\ 2007, in Protostars and Planets V,
eds.\ B.~Reipurth, D.~Jewitt, \& K.~Keil, (Tucson: University of Arizona Press), 81

\bibitem[Bohlin, Savage \& Drake(1978)]{bsd78}
Bohlin, R.~C., Savage, B.~D., Drake, J.~F.\ 1978, ApJ, 224, 132

\bibitem[Bonnell et al.(2006)]{bonnell06} Bonnell, I.~A., Dobbs, 
C.~L., Robitaille, T.~P., \& Pringle, J.~E.\ 2006, MNRAS, 365, 37 

%\bibitem[Burton, Hollenbach \& Tielens(1990)]{bht90}
%Burton, M.~G., Hollenbach, D.~J., \& Tielens, A.~G.~G.~M. 1990, ApJ,  365, 620

%\bibitem[Chabrier(2001)]{chabrier01}
%Chabrier, G.\ 2001, ApJ, 554, 1274

\bibitem[Clark et~al.(2011)]{clark11}
Clark, P.~C., Glover, S.~C.~O., Klessen, R.~S., \& Bromm, V.\ 2011, ApJ, 727, 110

%\bibitem[Clark, Glover \& Klessen(2008)]{clark08}
%Clark, P.~C., Glover, S.~C.~O.,  \& Klessen, R.~S.\ 2008, ApJ, 672, 757

\bibitem[Clark, Glover \& Klessen(2012)]{cgk12}
Clark, P.~C., Glover, S.~C.~O., \& Klessen, R.~S.\ 2012, MNRAS, 420, 745

\bibitem[Dame et al.(2001)]{Dame2001} Dame, T.~M., Hartmann, D., 
\& Thaddeus, P.\ 2001, ApJ, 547, 792 

\bibitem[van Dishoeck \& Black(1988)]{db88} van Dishoeck, E.~F., \& Black, J.~H.\ 1988, ApJ, 334, 771 

\bibitem[Dobbs \& Bonnell(2007)]{db07} Dobbs, C.~L., \& Bonnell, I.~A.\ 2007, MNRAS, 374, 1115

\bibitem[Dobbs et~al.(2008)]{dobbs08}
Dobbs, C. L., Glover, S.~C.~O., Clark, P.~C., \& Klessen, R.~S.\ 2008, MNRAS, 389, 1097

\bibitem[Dobbs(2008)]{dobbs2008} Dobbs, C.~L.\ 2008, MNRAS, 391, 
844

%\bibitem[Dopcke et al.(2011)]{dopcke11}
%Dopcke, G., Glover, S.~C.~O., Clark, P.~C., \& Klessen, R.~S.\ 2011, ApJ, 729, L3

\bibitem[Draine(1978)]{dr78}
Draine, B.~T. 1978, ApJS,  36, 595

\bibitem[Draine \& Bertoldi(1996)]{db96}
Draine, B.~T., \& Bertoldi, F.\ 1996, ApJ, 468, 269

\bibitem[Elmegreen(1993)]{elmegreen93}
Elmegreen, B.~G.\ 1993, ApJ, 419, L29

\bibitem[Elmegreen(2000)]{elmegreen00}
Elmegreen, B.~G.\ 2000, ApJ, 530, 277

\bibitem[Elmegreen(2007)]{elmegreen07} 
Elmegreen, B.~G.\ 2007, ApJ, 668, 1064

\bibitem[Elmegreen et al.(2008)]{elmegreen2008} 
Elmegreen, B.~G., Klessen, R.~S., \& Wilson, C.~D.\ 2008, ApJ, 681, 365 

\bibitem[Ferri\`{e}re(2001)]{fer01}
Ferri\`{e}re. K.~M.\ 2001, Rev.\ Mod.\ Phys., 73, 1031

\bibitem[Field(1965)]{field65}
Field, G.~B.\ 1965, ApJ, 142, 531

\bibitem[Glover \& Clark(2012a)]{gc12a}
Glover, S.~C.~O., \& Clark, P.~C.\ 2012a, MNRAS, 421, 9

\bibitem[Glover \& Clark(2012b)]{gc12b}
Glover, S.~C.~O., \& Clark, P.~C.\ 2012b, MNRAS, 421, 116

\bibitem[Glover et al.(2010)]{g10}
Glover, S.~C.~O., Federrath, C., {Mac Low}, M.-M., \& Klessen, R.~S.\ 2010, MNRAS, 404, 2

%\bibitem[Glover \& Jappsen(2007)]{gj07}
%Glover, S.~C.~O., \& Jappsen, A.-K.\ 2007, ApJ, 666, 1

\bibitem[Glover \& {Mac Low}(2007a)]{gm07a}
Glover, S.~C.~O., \& {Mac Low}, M.-M.\ 2007, ApJS, 169, 239

\bibitem[Glover \& {Mac Low}(2007b)]{gm07b}
Glover, S.~C.~O., \& {Mac Low}, M.-M.\ 2007, ApJ, 659, 1317

\bibitem[Glover \& {Mac Low}(2011)]{gm11}
Glover, S.~C.~O., \& {Mac Low}, M.-M.\ 2011, MNRAS, 412, 337

\bibitem[Grenier et~al.(2005)]{gren05}
Grenier, I.~A., Casandjian, J.-M., \& Terrier, R.\ 2005, Science, 307, 1292

\bibitem[Goldsmith(2001)]{gold01}
Goldsmith, P.~F. 2001, ApJ, 557, 736

\bibitem[Goldsmith et al.(2008)]{goldsmith2008} Goldsmith, P.~F., 
Heyer, M., Narayanan, G., et al.\ 2008, ApJ, 680, 428

\bibitem[G\'orski et~al.(2005)]{healpix}
G\'orski, K.~M., Hivon, E., Banday, A.~J., Wandelt, B.~D., Hansen, F.~K., Reinecke, M., 
\& Bartelmann, M.\  2005, ApJ, 622, 759

%\bibitem[Gnedin \& Kravtsov(2011)]{gk11}
%Gnedin, N.~Y., \& Kravtsov, A.~V.\ 2011, ApJ, 728, 88

\bibitem[Hartmann et~al.(2001)]{hartmann01}
Hartmann, L., Ballesteros-Paredes, J., Bergin, E.~A.\ 2001, ApJ, 562, 852

\bibitem[Heitsch et al.(2006)]{Heitsch2006} Heitsch, F., Slyz, 
A.~D., Devriendt, J.~E.~G., Hartmann, L.~W., 
\& Burkert, A.\ 2006, ApJ, 648, 1052 

\bibitem[Heitsch \& Hartmann(2008)]{hh08}
Heitsch, F., \& Hartmann, L.\ 2008, ApJ, 689, 290

\bibitem[Heitsch, Hartmann \& Burkert(2008)]{hhb2008} 
Heitsch, F., Hartmann, L.~W., \& Burkert, A.\ 2008a, ApJ, 683, 786

\bibitem[Heitsch et al.(2008)]{h08}
Heitsch, F., Hartmann, L.~W., Slyz, A.~D., Devriendt, J.~E.~G., \& Burkert, A.\ 2008, ApJ, 674, 316

\bibitem[Hennebelle \& P\'erault(1999)]{hp99}
Hennebelle, P., P\'erault, M., 1999, A\&A, 351, 309

\bibitem[Hennebelle \& P\'erault(2000)]{hp00}
Hennebelle, P., P\'erault, M., 2000, A\&A, 359, 1124

\bibitem[Hennebelle et al.(2007)]{Hennebelle2007} Hennebelle, P., Audit, E., \& Miville-Desch{\^e}nes, M.-A.\ 2007, A\&A, 465, 445 

\bibitem[Hennebelle et al.(2008)]{Hennebelle2008} Hennebelle, P., Banerjee, R., V{\'a}zquez-Semadeni, E., Klessen, R.~S., \& Audit, E.\ 2008, A\&A, 486, L43

\bibitem[Hollenbach \& McKee(1979)]{hm79} Hollenbach, D., \& McKee, C.~F.\ 1979, ApJS, 41, 555

%\bibitem[Hubber et al.(2006)]{hgw06} 
%Hubber, D.~A., Goodwin, S.~P., \& Whitworth, A.~P.\ 2006, A\&A, 450, 881 

\bibitem[Heyer \& Brunt(2004)]{hb2004} Heyer, M.~H., \& Brunt, C.~M.\ 2004, ApJL, 615, L45

\bibitem[Heyer et al.(2009)]{heyer2009} Heyer, M., Krawczyk, C., 
Duval, J., \& Jackson, J.~M.\ 2009, ApJ, 699, 1092 

\bibitem[Jappsen et~al.(2005)]{jappsen05}
Jappsen, A.-K., Klessen, R.~S., Larson, R.~B., Li, Y., {Mac Low}, M.-M. 2005,
A\&A, 435, 611

\bibitem[Jeffries et al.(2011)]{Jeffries2011} Jeffries, R.~D., 
Littlefair, S.~P., Naylor, T., \& Mayne, N.~J.\ 2011, MNRAS, 418, 1948 

%\bibitem[Komugi et~al.(2011)]{komugi11}
%Komugi, S., Tasui, C., Kobayashi, N., Hatsukade, B., Kohno, K., Sofue, Y., \& Kyu, S.\ 2011, PASJ, 63, L1

\bibitem[Koyama \& Inutsuka(2000)]{ki00}
Koyama, H., \& Inutsuka, S.\ 2000, ApJ, 532, 980

\bibitem[Koyama \& Inutsuka(2002)]{ki02}
Koyama, H., \& Inutsuka, S.\ 2002, ApJ,  564, L97

%\bibitem[Kroupa(2002)]{kroupa02}
%Kroupa, P.\ 2002, Science, 295, 82

\bibitem[Krumholz, Leroy \& McKee(2011)]{klm11}
Krumholz, M.~R., Leroy, A.~K., \& McKee, C.~F.\ 2011, ApJ, 731, 25

%\bibitem[Krumholz \& McKee(2005)]{krumholzmckee05} 
%Krumholz, M.~R., \& McKee, C.~F.\ 2005, ApJ, 630, 250

%\bibitem[Krumholz et al.(2009)]{kmt09}
%Krumholz, M.~R., McKee, C.~F., \& Tumlinson, J.\ 2009, ApJ, 699, 850

\bibitem[Krumholz \& McKee(2005)]{km05}
Krumholz, M.~R., McKee, C.~F.\ 2005, ApJ, 630, 250

\bibitem[Langer et~al.(2010)]{langer10}
Langer, W.~D., Velusamy, T., Pineda, J.~L., Goldsmith, P.~F., Li, D., \& Yorke, H.~W.\ 2010, A\&A, 521, L17

\bibitem[Larson(1985)]{larson85}
Larson, R.\ 1985, MNRAS, 214, 379

\bibitem[Larson(2005)]{larson05}
Larson, R.\ 2005, MNRAS, 359, 211

\bibitem[Lee et~al.(1996)]{lee96}
Lee, H.-H., Herbst, E., {Pineau des For\^ets}, G.,
Roueff, E., \& {Le Bourlot}, J.\ 1996, A\&A, 311, 690

%\bibitem[Leroy et al.(2007)]{leroy07}
%Leroy, A.,  Cannon, J., Walter, F., Bolatto, A., \& Weiss, A.\ 2007, ApJ, 663, 990

%\bibitem[Leroy et al.(2008)]{leroy08} 
%Leroy, A.~K., Walter, F.,  Brinks, E., Bigiel, F., de Blok, W.~J.~G., Madore, B., 
%\& Thornley, M.~D.\ 2008, AJ, 136, 2782 

\bibitem[{Le Teuff}, Millar \& Markwick(2000)]{teu00}
{Le Teuff}, Y.~H., Millar, T.~J., \& Markwick, A.~J. 2000, A\&AS, 146, 157

%\bibitem[Li, Klessen, \& {Mac Low}(2003)]{li03}
%Li, Y., Klessen, R.~S., \& {Mac Low}, M.-M.\ 2003, ApJ, 592, 975

\bibitem[Mac Low \& Glover(2012)]{mg12}
{Mac Low}, M.-M., \& Glover, S.~C.~O.\ 2012, ApJ, 746, 135

\bibitem[Mac Low 
\& Klessen(2004)]{mlk2004} Mac Low, M.-M., \& Klessen, R.~S.\ 2004, Reviews of Modern Physics, 76, 125 

\bibitem[Mathis, Mezger \& Panagia(1983)]{mmp83} 
Mathis, J.~S., Mezger, P.~G., \& Panagia, N. 1983,  A\&A, 128, 212

\bibitem[McKee 
\& Ostriker(2007)]{mo2007} McKee, C.~F., \& Ostriker, E.~C.\ 2007, ARA\&A, 45, 565

\bibitem[Micic et~al.(2012)]{micic12}
Micic, M., Glover, S.~C.~O., Federrath, C., Klessen, R.~S.\ 2012, MNRAS, 421, 2531

\bibitem[Molina et~al.(2011)]{molina11}
Molina, F., Glover, S., \& Federrath, C.\ 2011, in ``Conditions and Impact of Star Formation: 
New Results with Herschel and Beyond'',  EAS Publications Series 52, 
eds.\ M.~R{\"o}llig, R.~Simon, V.~Ossenkopf, \& J.~Stutzki, 289

\bibitem[Mouschovias, Tassis \& Kunz(2006)]{mtk06}
Mouschovias, T.~Ch., Tassis, K., \& Kunz, M.~W.\ 2006, ApJ, 646, 1043

\bibitem[Narayanan et al.(2008)]{narayanan2008} Narayanan, G., Heyer, 
M.~H., Brunt, C., et al.\ 2008, ApJS, 177, 341 

\bibitem[Nelson \& Langer(1999)]{nl99}
Nelson, R.~P., \& Langer, W.~D. 1999, ApJ, 524, 923

%\bibitem[Neufeld, Lepp \& Melnick(1995)]{nlm95}
%Neufeld, D.~A., Lepp., S., \& Melnick, G.~J. 1995, ApJS, 100, 132

\bibitem[Ossenkopf \& Henning(1994)]{oh94} Ossenkopf, V., \& Henning, Th. 1994, A\&A, 291, 943

%\bibitem[Padoan \& Nordlund(2011)]{pn11}
%Padoan, P., \& Nordlund, \AA.\ 2011, ApJ, 730, 40

\bibitem[Pagani, Roueff \& Lesaffre(2011)]{pagani11}
Pagani, L., Roueff, E., \& Lesaffre, P.\ 2011, ApJ, 739, L35

%\bibitem[Pan \& Padoan(2009)]{pp09}
%Pan, L., \& Padoan, P.\ 2009, ApJ, 692, 594

\bibitem[Pineda et~al.(2010)]{pineda10}
Pineda, J.~L., Velusamy, T., Langer, W.~D., Goldsmith, P.~F., Li, D., \& Yorke, H.~W.\ 2010, A\&A, 521, L19

\bibitem[Pringle, Allen \& Lubow(2001)]{pal01}
Pringle, J.~E., Allen, R.~J., \& Lubow, S.~H.\ 2001, MNRAS, 327, 663

\bibitem[R\"ollig et~al.(2007)]{roellig07}
R\"ollig, M., et~al., 2007, A\&A, 467, 187

\bibitem[Rosas-Guevara et~al.(2010)]{rg10}
Rosas-Guevara, Y., V\'{a}zquez-Semadeni, E., G\'{o}mez, G.~C., Jappsen, A.-K.\ 2010, MNRAS, 406, 1875

%\bibitem[Schaye(2004)]{schaye04} 
%Schaye, J.\ 2004, ApJ, 609, 667

\bibitem[Sembach et~al.(2000)]{sem00}
Sembach, K.~R., Howk, J.~C., Ryans, R.~S.~I., \& Keenan, F.~P.
2000, ApJ,  528, 310

\bibitem[Shetty et al.(2011a)]{Shetty2011a} 
Shetty, R., Glover, S.~C., Dullemond, C.~P., \& Klessen, R.~S.\ 2011a, MNRAS, 412, 1686

\bibitem[Shetty et al.(2011b)]{Shetty2011b} 
Shetty, R., Glover, S.~C., Dullemond, C.~P.,  Ostriker, E.~C., Harris, A.I., Klessen, R.~S.\ 2011b, MNRAS, 415, 3253

\bibitem[Sobolev(1957)]{Sobolev57} Sobolev, V.~V.\ 1957, SvA, 1678

\bibitem[Solomon et al.(1987)]{Solomon1987} Solomon, P.~M., Rivolo, 
A.~R., Barrett, J., \& Yahil, A.\ 1987, ApJ, 319, 730

\bibitem[Springel(2005)]{springel05}
Springel, V.\ 2005, MNRAS, 364, 1105

%\bibitem[Takahashi \& Uehara(2001)]{tu01}
%Takahashi, J., \& Uehara, H.\ 2001, ApJ, 561, 843

\bibitem[Tamburro et~al.(2008)]{tamburro08}
Tamburro, D., Rix, H.-W., Walter, F., Brinks, E., {de Blok}, W.~J.~G., Kennicutt, R.~C., \& {Mac Low}, M.-M.\ 2008, AJ, 136, 2872 

%\bibitem[Tasker \& Tan(2009)]{tt09}
%Tasker, E.~J., \& Tan, J.~C.\ 2009, ApJ, 700, 358

\bibitem[Tassis \& Mouschovias(2004)]{tassis04}
Tassis, K., \& Mouschovias, T.~Ch., 2004, ApJ, 616, 283

\bibitem[V{\'a}zquez-Semadeni et al.(2006)]{enrique2006} 
V{\'a}zquez-Semadeni, E., Ryu, D., Passot, T., Gonz{\'a}lez, R.~F., 
\& Gazol, A.\ 2006, ApJ, 643, 245 

\bibitem[V\'{a}zquez-Semadeni et al.(2007)]{vs07}
V\'{a}zquez-Semadeni, E., G\'omez, G., Jappsen, A.-K., Ballesteros-Paredes, J., 
Gonz\'{a}lez, R.~F., \& Klessen, R.~S.\ 2007, ApJ, 657, 870

\bibitem[V{\'a}zquez-Semadeni et al.(2009)]{enrique2009} 
V{\'a}zquez-Semadeni, E., G{\'o}mez, G.~C., Jappsen, A.-K., 
Ballesteros-Paredes, J., \& Klessen, R.~S.\ 2009, ApJ, 707, 1023 

\bibitem[Velusamy et~al.(2010)]{velu10}
Velusamy, T., Langer, W.~D., Pineda, J.~L., Goldsmith, P.~F., Li, D., \& Yorke, H.~W.\ 2010, A\&A, 521, L18

\bibitem[Visser, {van Dishoeck}, \& Black(2009)]{visser09}
Visser, R., {van Dishoeck}, E.~F., \& Black, J.~H.\ 2009, A\&A, 503, 323

\bibitem[Walder \& Folini(1998a)]{WalderFolini1998a} Walder, R., \& Folini, D.\ 1998, Ap\&SS , 260, 215 

\bibitem[Walder \& Folini(1998b)]{WalderFolini1998b} Walder, R., \& Folini, D.\ 1998, A\&A, 330, L21 

%\bibitem[Whitworth(1998)]{w98} Whitworth, A.~P.\ 1998, MNRAS, 296, 442

\bibitem[Wolfire et al.(1995)]{wolf95}
Wolfire, M.~G., Hollenbach, D., McKee, C.~F., Tielens, A.~G.~G.~M., \& Bakes, E.~L.~O.\ 1995,
ApJ, 443, 152

\bibitem[Wolfire et al.(2003)]{wolf03}
Wolfire, M.~G., McKee, C.~F., Hollenbach, D., \& Tielens, A.~G.~G.~M.\ 2003,
ApJ, 587, 278

\bibitem[Wolfire et al.(2010)]{Wolfire2010} Wolfire, M.~G., 
Hollenbach, D., \& McKee, C.~F.\ 2010, ApJ, 716, 1191

%\bibitem[Wong \& Blitz(2002)]{wong02} 
%Wong, T., \& Blitz, L.\ 2002, ApJ, 569, 157

\bibitem[Young \& Scoville(1991)]{Young1991} Young, J.~S., \& Scoville, N.~Z.\ 1991, ARA\&A, 29, 581

\end{thebibliography}
\end{document}